\documentclass{article}
\usepackage{ijcai26}

\usepackage{times}
\usepackage{soul}
\usepackage{url}
\usepackage[hidelinks]{hyperref}
\usepackage[utf8]{inputenc}
\usepackage[small]{caption}
\usepackage{graphicx}
\usepackage{amsmath}
\usepackage{amsthm}
\usepackage{booktabs}
\usepackage{algorithm}
\usepackage{algorithmic}
\usepackage[switch]{lineno}
\usepackage{hyperref}
\usepackage{cleveref}
\usepackage{amsfonts}
\usepackage{bm}
\usepackage{multirow}
\usepackage[normalem]{ulem}
\useunder{\uline}{\ul}{}
\usepackage{bbding}
\usepackage{makecell}
\usepackage{xcolor}
\title{TME-PSR: Time-aware, Multi-interest, and Explanation Personalization for Sequential Recommendation}

\author{
Qingzhuo Wang$^1$
\and
Leilei Wen$^1$\and
Juntao Chen$^{1}$\and
Kunyu Peng$^{1}$\and
Ruiyang Qin$^1$\and
Zhihua Wei$^1$\and
Wen Shen$^1$\\
\affiliations
$^1$Tongji University\\
\emails
2534123@tongji.edu.cn,
wenleilei11@gmail.com,
\{2432047, 2534000\}@tongji.edu.cn,
wyattqin@gmail.com,
\{zhihua\_wei, wenshen\}@tongji.edu.cn,
}

\begin{document}

\maketitle

\begin{abstract}
In this paper, we propose a sequential recommendation model that integrates {\underline {\bf T}}ime-aware personalization, {\underline {\bf M}}ulti-interest personalization, and {\underline {\bf E}}xplanation personalization for {\underline {\bf P}}ersonalized {\underline {\bf S}}equential {\underline {\bf R}}ecommendation ({\bf TME-PSR}). That is, we consider the differences across different users in temporal rhythm preference,  multiple fine-grained latent interests, and the personalized semantic alignment between recommendations and explanations. Specifically, the proposed TME-PSR model employs a dual-view gated time encoder to capture personalized temporal rhythms, a lightweight multihead Linear Recurrent Unit architecture that enables fine-grained sub-interest modeling with improved efficiency, and a dynamic dual-branch mutual information weighting mechanism to achieve personalized alignment between recommendations and explanations. Extensive experiments on real-world datasets demonstrate that our method consistently improves recommendation accuracy and explanation quality, at a lower computational cost.
\end{abstract}

\section{Introduction}

In recent years, personalized recommendation has attracted increasing attention as user expectations rise in recommendation systems. Personalized recommendation aims to select items (information or products) that best suit individual users' preferences based on their historical behaviors. Although recent advances in sequential recommendation (SR) models have significantly improved recommendation performance, most methods still fall short in delivering comprehensive personalization. 
In this paper, we aim to model three types of personalization simultaneously, including the following three perspectives.

\begin{figure}[tbp]
  \centering
  \includegraphics[width=1\linewidth]{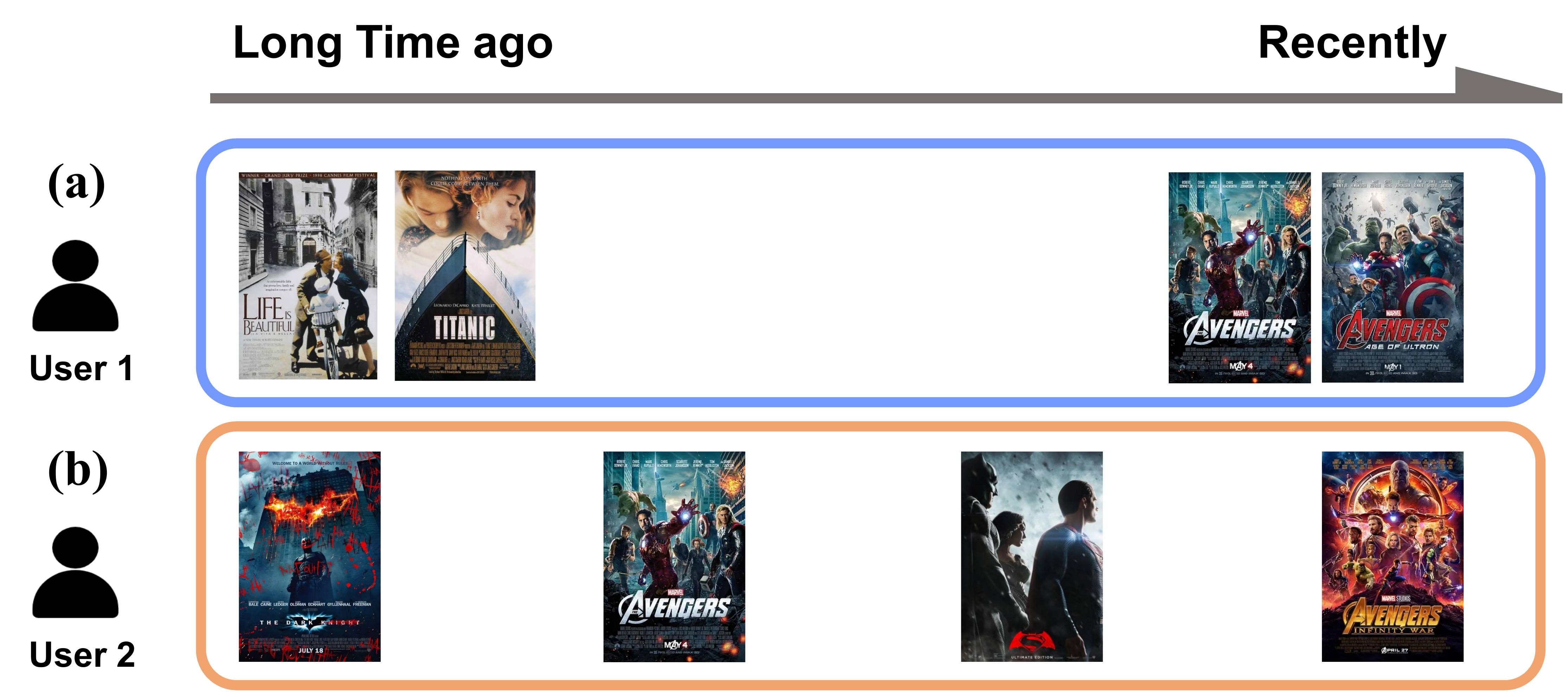}
  \caption{Examples of movie interaction sequences of two different users, which show different preferences for time and interests. User 1 liked romance movies a long time ago, but recently the interest has shifted to superhero movies. User 2 has always liked superhero movies, but also follows both Marvel series and DC series two sub-interests.} 
  \label{fig:three_personalization}
\end{figure}

$\bullet$ {\bf Temporal rhythm preference}. Given a set of historical interactions of a specific user, the interaction time can reflect the personalized temporal rhythm preference of the user.  For example, as shown in \cref{fig:three_personalization}, User 1 watches two romantic dramas on one day, and then, after a long gap of nearly a month, suddenly watches two action movies, which suggests a clear shift in temporal rhythm. In contrast, User 2 watches superhero movies regularly. 
To encode such long-short-term temporal rhythms across different users, some studies design specific network architectures \cite{slirec,tlsan,clsr,TiCoSeRec,tisasrec,tlrec,unirec,rtisr,mojito} to improve the recommendation performance, such as using two attention branches to encode long-term and short-term temporal rhythms, respectively. In comparison, we propose a {\it flexible module to adaptively encode the differences in long-short-term temporal rhythms among different users, which can be plugged into existing models without changing the model's original architecture}. More crucially, the proposed module is not limited to recommendation tasks, it also encodes temporal rhythm preferences for the explanation task, separately for the same user.

$\bullet$ \textbf{Multiple fine-grained latent interests}. A single sequence may contain multiple latent interests of a user, such as the interest in superhero movies and the interest in romance movies. 
Moreover, a user's latent interest can be further divided into {\bf multiple fine-grained sub-interests}, such as the sub-interest in DC series movies and the sub-interest in Marvel series movies, as \cref{fig:three_personalization}(b) shows. 
Existing studies introduce an additional attention layer or an additional capsule network \cite{mindrec,pimirec,comirec,re4,umi,remi,teddy} to learn multiple interests for different users. However, the time and space complexity of these methods increases with the number of multiple interests. In comparison, we propose a method to {\it make the model automatically learn multiple fine-grained sub-interests with parallel optimized network branches}, thereby explicitly disentangling them while improving efficiency and parameter economy.

$\bullet$ {\bf Personalized semantic alignment between recommendations and explanations}. As explainability gains importance in recommendation systems, more users desire a textual rationale explaining why the item is recommended to them. Many studies, such as collaborative explanation \cite{dataset,bper,scemim}, fuse recommendation features and explanation features to achieve semantic alignment between recommendations and explanations, thereby enabling explainable recommendation. However, these methods assume that the degree of semantic alignment between recommendations and explanations is uniform across users. 
But different users have different requirements for the alignment degree. For example, if a user often watches Marvel's superhero movies in the past, giving an explanation such as {\it ``Iron Man 3, Robert Downey Jr. makes a powerful comeback''} can effectively entice the user to choose this movie, this user has high requirements for the semantic alignment degree between recommendations and explanations. In contrast, for those users having a wide range of interests, this explanation might be less appealing, as they may not be familiar with {\it``Iron Man''} or {\it``Robert Downey Jr.''} In this case, generating an explanation like {\it``An excellent action movie''} might be more likely to entice them to watch. These users have low requirements for the semantic alignment degree. Therefore, it is necessary to consider differences in the degree of semantic alignment between recommendations and explanations among different users, which is termed {\it explanation personalization}.

\begin{figure*}[tbp]
  \centering
  \includegraphics[width=1\textwidth]{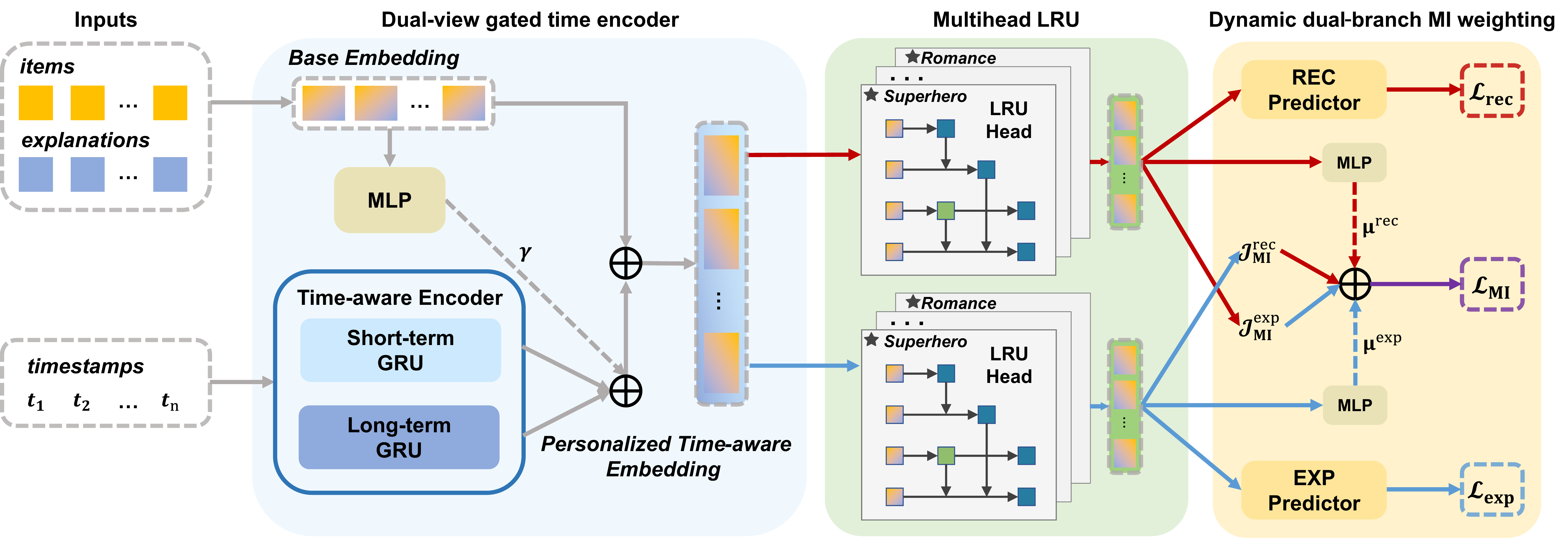}
  \caption{Overall architecture of our TME-PSR. At the input layer, we propose a dual-view gated time encoder. Specifically, we use a time-aware encoder to extract short-term behavior patterns and long-term behavior patterns from input timestamps. Then, we adaptively weight the two patterns based on different user’s temporal rhythm preference, so as to extract personalized time-aware embeddings. At the encoding layer, we propose a multihead LRU architecture, each head captures a fine-grained sub-interest. At the prediction layer, we design a dynamic dual-branch MI weighting mechanism, so as to adaptively adjust the degree of semantic alignment between the recommendation branch and the explanation branch.}
  \label{fig:overall}
\end{figure*}

To this end, in this paper, we propose a sequential recommendation model to achieve the above three types of personalization. Specifically, as shown in \cref{fig:overall}, the overall structure of our model is given as follows:

$\bullet$ To realize the time-aware personalization, we propose a {\it dual-view gated time encoder module}. First, the module adaptively balances short-term and long-term behavior patterns based on each user’s temporal rhythm preference. Second, for the same user, the module independently considers the temporal rhythm preference of the recommendation task and the explanation task. The proposed module is plug-and-play, requiring no changes to existing backbones.

$\bullet$ To realize the multi-interest personalization, we propose to divide the representations of intermediate layers into multiple subspaces and track independent sub-interests in each subspace. Specifically, we extend the Linear Recurrent Unit (LRU) \cite{lru} to a {\it multihead LRU architecture}, so as to disentangle multiple fine-grained latent sub-interests of a certain user while keeping lightweight. 

$\bullet$ To realize the explanation personalization, we design a {\it dynamic dual-branch mutual information (MI) weighting mechanism}, inspired by \cite{scemim}, so as to adaptively adjust the degree of semantic alignment between recommendations and explanations.

\section{RELATED WORK}

{\bf Time-Aware Sequential Recommendation.} Time-aware methods studies incorporate temporal information to improve recommendation performance. Early works utilized time matrix factorization \cite{fmtemporal,dingtime,xiongtemporal} or attention mechanisms \cite{tisasrec,tlsan,tlrec} to encode time intervals. Later, some methods \cite{TiCoSeRec,mojito,unirec} have improved model performance by enhancing the uniformity of data through time intervals to obtain more regular historical sequences. Recent methods also study temporal phenomena such as repeat consumption \cite{recode}, temporal graph contrastive learning \cite{tgcl} or review-driven \cite{rtisr}. Above methods all propose a specific model structure, but many require architecture-coupled designs, which limits portability.

{\bf Multi-Interest Sequential Recommendation.} To model users’ multifaceted preferences, many methods learn multiple user vectors. Capsule/attention aggregators \cite{mindrec,comirec} and their variants \cite{pimirec,umi} extract several interests for matching/ranking. Recent methods \cite{MBTGT,re4,remi,dmrl,teddy} have improved the model's ability to decompose multiple interests and robustness. However, complexity often scales with interest count, leading to overlapping/redundancy of interests and higher computational costs.


{\bf Explainable Sequential Recommendation.} Explainable SR provides textual rationales. Approaches utilize attention weights \cite{attnexpl1,attnexpl2,attnexpl3,attnexpl4}, knowledge graphs \cite{pathexpl1,pathexpl2,pathexpl3,pathexpl4}, or collaborative generation \cite{dataset,bper,scemim}. But existing collaborative explanation methods typically assume a fixed alignment degree between recommendations and explanations across users.

Broader advances in general sequential recommendation are discussed in App. A.

\section{METHODOLOGY}

{\bf Problem Formulation.} Let $\mathcal{U}$ denote the set of users, $\mathcal{V}$ denote the set of items, and $\mathcal{E}$ denote the set of explanations.\footnote{The explanations are extracted from user reviews and clustered into representative phrases through semantic hashing techniques \cite{dataset}, where each item is assigned a unique explanation.} For each user $u\in \mathcal{U}$, let $\mathcal{S}_u = \left[ (v_1, e_1, t_1), (v_2, e_2, t_2), \dots, (v_n, e_n, t_n) \right]$ denote the interaction sequence of the user $u$, where at time step $i\in \{1,2,...,n\}$, the user action is a tuple of $(v_i,e_i,t_i)$,\ $v_i \in \mathcal{V}$ is the item interacted at time step $i$, $e_i \in \mathcal{E}$ is the associated explanation, and $t_i \in \mathbb{R}^+$ is the interaction timestamp. We propose a dual-branch model TME-PSR, as \cref{fig:overall} shows, one branch predicts the next item $v_{i+1}$ that user $u$ is likely to interact with, while the other generates the corresponding explanation $e_{i+1}$ for this item. The proposed TME-PSR model achieves time-aware personalization, multi-interest personalization, and explanation personalization, as follows.

\subsection{Time-Aware Personalization}\label{sec:timeaware}
To model user-specific temporal rhythm preference, we propose a {\it dual-view gated time encoder} to extract the personalized time-aware embedding from the timestamps of the user sequence $S_u$. Specifically, we first encode two types of time intervals: the interval between adjacent items and the interval between each item and the first item in the sequence, thereby obtaining features containing short-term and long-term temporal information, respectively. Then, we employ a gating mechanism to adaptively fuse the short-term features and the long-term features, producing the personalized time-aware embedding of the user sequence $S_u$.

{\bf Base embedding construction.} Let $\mathbf{M}_\mathcal{V} \in \mathbb{R}^{|\mathcal{V}| \times d}$ and $\mathbf{M}_\mathcal{E} \in \mathbb{R}^{|\mathcal{E}| \times d}$ denote the embedding matrices for all items $\mathcal{V}$ and all explanations $\mathcal{E}$, respectively, where $d$ is the embedding size. Given a user sequence $\mathcal{S}_u$ with $n$ interactions, we first apply a look-up operation from $\mathbf{M}_\mathcal{V}$ and $\mathbf{M}_\mathcal{E}$ to form the user's item embedding matrix $\mathbf{E}_\mathcal{V} \in \mathbb{R}^{n \times d}$ and explanation embedding matrix $\mathbf{E}_\mathcal{E} \in \mathbb{R}^{n \times d}$. We construct the recommendation base embeddings\footnote{\label{ft:2}The embeddings (e.g. $\mathbf{E}^{\text{rec}}$) and hidden states extracted in the METHODOLOGY section are all for a single user sequence $S_u$. The subscript $u$ is omitted in these symbols for simplicity.} $\mathbf{E}^{\text{rec}}\in \mathbb{R}^{n \times d}$ and the explanation base embeddings\textsuperscript{\ref{ft:2}} $\mathbf{E}^{\text{exp}}\in \mathbb{R}^{n \times d}$ as follows, which will serve as a basis for constructing personalized time-aware embedding:
\begin{small}
\begin{equation}
\mathbf{E}^{\text{rec}} = \alpha \cdot \mathbf{E}_\mathcal{V} + (1 - \alpha) \cdot \mathbf{E}_\mathcal{E}, \quad \mathbf{E}^{\text{exp}} = \alpha \cdot \mathbf{E}_\mathcal{E} + (1 - \alpha) \cdot \mathbf{E}_\mathcal{V},
\end{equation}
\end{small}
where $\alpha \in [0,1]$ is a hyperparameter that controls the relative importance of each item embedding and its corresponding explanation embedding.

{\bf Dual-view gated time encoder.} Let $\mathcal{T} = [t_1, t_2, ..., t_n] \in \mathbb{R}^n$ denote the timestamp sequence corresponding to the interaction sequence of the user $u$\textsuperscript{\ref{ft:2}}, we compute two types of time interval to extract short-term and long-term temporal features as follows: 

(1) We compute the difference between adjacent timestamps $t_i$ and $t_{i-1}$ to obtain the {\it adjacent time interval} $\bm{\Delta}^{\text{adj}}_i = \log(1 + t_i - t_{i-1}) \in \mathbb{R}^+,\ \text{for} \ i=2,3,...,n$, where the $\log$ operation is to ensure numerical smoothness; the constant 1 is added to avoid taking the $\log$ of zero. 

(2) We compute the difference between each timestamp $t_i$ and the first timestamp $t_{1}$ to obtain the {\it absolute time interval} $\bm{\Delta}^{\text{abs}}_i = \log(1 + t_i - t_1) \in \mathbb{R}^+,\ \text{for} \ i=2,3,...,n$.

Therefore, we construct the adjacent interval vector $\bm{\Delta}^{\text{adj}}=[0,\Delta^{\text{adj}}_2,...,\Delta^{\text{adj}}_n]\in \mathbb{R}^n$ and the absolute interval vector $\bm{\Delta}^{\text{abs}}=[0,\Delta^{\text{abs}}_2,...,\Delta^{\text{abs}}_n]\in \mathbb{R}^n$, where the constant 0 is prepended to each vector to align with the length $n$ of the user sequence $S_u$ to facilitate subsequent matrix operations.

To extract short-term temporal features from $\bm{\Delta}^{\text{adj}}$ and extract long-term temporal features from $\bm{\Delta}^{\text{abs}}$, we employ two shared GRU \cite{gru} networks, $\mathrm{GRU}_{\text{adj}}(\cdot): \mathbb{R}^n\rightarrow \mathbb{R}^{n\times d}$ and $\mathrm{GRU}_{\text{abs}}(\cdot): \mathbb{R}^n\rightarrow \mathbb{R}^{n \times d}$, as time encoding networks:
\begin{small}
\begin{equation}
\mathbf{H}^{\text{adj}} = \mathrm{GRU}_{\text{adj}}(\bm{\Delta}^{\text{adj}}), \quad \mathbf{H}^{\text{abs}} = \mathrm{GRU}_{\text{abs}}(\bm{\Delta}^{\text{abs}}),
\end{equation}
\end{small}
where features\textsuperscript{\ref{ft:2}} $\mathbf{H}^{\text{adj}} \in \mathbb{R}^{n \times d}$ represent the short-term temporal rhythm and features\textsuperscript{\ref{ft:2}} $\mathbf{H}^{\text{abs}} \in \mathbb{R}^{n \times d}$ represent the long-term temporal rhythm. 

To learn personalized time-aware embeddings, we propose to learn two gating parameters $\gamma^{\text{rec}}$ and $\gamma^{\text{exp}}$ for recommendation embeddings and explanation embeddings, respectively. These gates adaptively fuse the two features:
\begin{small}
\begin{equation}
\label{eq:3}
\tilde{\mathbf{E}}^{\text{rec}} = \gamma^{\text{rec}} \cdot \mathbf{H}^{\text{adj}} + (1 - \gamma^{\text{rec}}) \cdot \mathbf{H}^{\text{abs}}, 
\end{equation}
\begin{equation}
\label{eq:4}
\tilde{\mathbf{E}}^{\text{exp}} = \gamma^{\text{exp}} \cdot \mathbf{H}^{\text{adj}} + (1 - \gamma^{\text{exp}}) \cdot \mathbf{H}^{\text{abs}},
\end{equation}
\end{small}
where the gating parameters $\gamma^{\text{rec}}$ and $\gamma^{\text{exp}}$ are learned as follows:
\begin{small}
\begin{equation}
\gamma^{\text{rec}} = \sigma(\mathrm{MLP}_{\text{rec}}(\bar{\mathbf{E}}^{\text{rec}})),\quad \gamma^{\text{exp}} = \sigma(\mathrm{MLP}_{\text{exp}}(\bar{\mathbf{E}}^{\text{exp}})),
\end{equation}
\end{small}
where $\mathrm{MLP}_{\text{rec}}(\cdot):\mathbb{R}^{d}\rightarrow \mathbb{R}^{1}$ and $\mathrm{MLP}_{\text{exp}}(\cdot):\mathbb{R}^{d}\rightarrow \mathbb{R}^{1}$ are two-layer perceptrons; $\bar{\mathbf{E}}^{\text{rec}}=\mathrm{Mean}1(\mathbf{E}^{\text{rec}})\in \mathbb{R}^{1\times d}$ and $\bar{\mathbf{E}}^{\text{exp}}=\mathrm{Mean}1(\mathbf{E}^{\text{exp}})\in \mathbb{R}^{1\times d}$; $\sigma$ is sigmoid activation.

Finally, the personalized time-aware embeddings are obtained by combining the base embeddings\textsuperscript{\ref{ft:2}} $\mathbf{E}^{\text{rec}}$ and $\mathbf{E}^{\text{exp}}$ and the temporal features\textsuperscript{\ref{ft:2}} $\tilde{\mathbf{E}}^{\text{rec}}$ and $\tilde{\mathbf{E}}^{\text{exp}}$:
\begin{small}
\begin{equation}
\mathbf{E}^{\text{rec-time}} = \mathbf{E}^{\text{rec}} + \beta \cdot \tilde{\mathbf{E}}^{\text{rec}}, \quad \mathbf{E}^{\text{exp-time}} = \mathbf{E}^{\text{exp}} + \beta \cdot \tilde{\mathbf{E}}^{\text{exp}},
\end{equation}
\end{small}
where $\beta$ is a scaling hyperparameter that controls the strength of temporal information.

\subsection{Multi-interest Personalization}\label{Multi-interest Personalization}

In this section, we further extract fine-grained sub-interests of users from the obtained personalized time-aware embeddings $\mathbf{E}^{\text{rec-time}}$ and $\mathbf{E}^{\text{exp-time}}$. To this end, we divide $\mathbf{E}^{\text{rec-time}}$ and $\mathbf{E}^{\text{exp-time}}$ into several sub-embeddings, and we extend LRU \cite{lru} to a multihead variant to extract sub-interests from these sub-embeddings. Specifically, given interaction $(v_i,e_i,t_i)$ of the $i$-th time step and its corresponding time-aware personalized recommendation embedding $\mathbf{E}_i^{\text{rec-time}}\in \mathbb{R}^{1\times d}$ and explanation embedding $\mathbf{E}_i^{\text{exp-time}}\in \mathbb{R}^{1\times d}$, we divide the recommendation embedding $\mathbf{E}_i^{\text{rec-time}}$ into $H$ sub-embeddings, $\mathbf{e}_i^{(\text{rec-time},1)},...,\mathbf{e}_i^{(\text{rec-time},H)}\in \mathbb{R}^{1\times \frac{d}{H}}$, and divide the explanation embedding $\mathbf{E}_i^{\text{exp-time}}$ into $H$ sub-embeddings, $\mathbf{e}_i^{(\text{exp-time},1)},...,\mathbf{e}_i^{(\text{exp-time},H)}\in \mathbb{R}^{1\times \frac{d}{H}}$. 

Then, we propose a multihead LRU architecture with $H$ independent recurrent streams and we apply the architecture to both the recommendation branch and the explanation branch. Let $\mathbf{W}^{\text{rec}}_h,\mathbf{U}^{\text{rec}}_h\in \mathbb{R}^{\frac{d}{H}\times \frac{d}{H}}$ and $\mathbf{W}^{\text{exp}}_h,\mathbf{U}^{\text{exp}}_h\in \mathbb{R}^{\frac{d}{H}\times \frac{d}{H}}$ denote the learnable matrices for the $h$-th head in the recommendation branch and explanation branch, respectively. Each head learns to focus on a specific sub-interest. The hidden state for each head at time step $i$ is computed as follows:
\begin{small}
\begin{equation}
\mathcal{H}_i^{(\text{rec},h)} = \sum\nolimits_{k=1}^i\mathbf{e}_k^{(\text{rec-time},h)}(\mathbf{W}^{\text{rec}}_h)^{i-k}\mathbf{U}^{\text{rec}}_h\in \mathbb{R}^{1\times \frac{d}{H}},
\end{equation}
\begin{equation}
\mathcal{H}_i^{(\text{exp},h)} = \sum\nolimits_{k=1}^i\mathbf{e}_k^{(\text{exp-time},h)}(\mathbf{W}^{\text{exp}}_h)^{i-k}\mathbf{U}^{\text{exp}}_h\in \mathbb{R}^{1\times \frac{d}{H}}.
\end{equation}
\end{small}
We stack the hidden states of all time steps to get the complete hidden state matrices $\mathcal{H}^{(\text{rec},h)}$ and $\mathcal{H}^{(\text{exp},h)}$ of the $h$-th head for recommendation branch and explanation branch, respectively, as follows:
\begin{small}
\begin{equation}
\mathcal{H}^{(\text{rec},h)}=\text{Stack}(\mathcal{H}_1^{(\text{rec},h)},\mathcal{H}_2^{(\text{rec},h)},...,\mathcal{H}_n^{(\text{rec},h)})\in \mathbb{R}^{n\times \frac{d}{H}},
\end{equation}
\begin{equation}
\mathcal{H}^{(\text{exp},h)}=\text{Stack}(\mathcal{H}_1^{(\text{exp},h)},\mathcal{H}_2^{(\text{exp},h)},...,\mathcal{H}_n^{(\text{exp},h)})\in \mathbb{R}^{n\times \frac{d}{H}}.
\end{equation}
\end{small}
Finally, the personalized multi-interest representations\textsuperscript{\ref{ft:2}} $\mathbf{Z}^\text{rec}$ for the recommendation branch and $\mathbf{Z}^\text{exp}$ for the explanation branch are obtained by concatenating all heads:
\begin{small}
\begin{equation}
\label{eq:11}
\mathbf{Z}^\text{rec} = \text{Concat}(\mathcal{H}^{(\text{rec},1)},\mathcal{H}^{(\text{rec},2)},...,\mathcal{H}^{(\text{rec},H)})\in \mathbb{R}^{n \times d},
\end{equation}
\begin{equation}
\label{eq:12}
\mathbf{Z}^\text{exp} = \text{Concat}(\mathcal{H}^{(\text{exp},1)},\mathcal{H}^{(\text{exp},2)},...,\mathcal{H}^{(\text{exp},H)})\in \mathbb{R}^{n \times d}.
\end{equation}
\end{small}

The proposed multihead LRU architecture not only captures multiple fine-grained sub-interests but also reduces the time and space complexity of the model. The time complexity of ideal parallel computing of all heads achieves $O(\log(n)\cdot(\frac{d}{H})^2)$. The space complexity achieves $O(\frac{d^2}{H})$. The per-step incremental inference complexity is $O(\frac{d^2}{H^2})$, independent of $n$. Please see App. F for the detailed complexity analysis. We demonstrate the advantages of the proposed multihead LRU on efficiency in \cref{fig:efficiency}.

\subsection{Explanation Personalization}\label{Explanation Personalization}

To measure the personalized degree of semantic alignment between recommendations and explanations, we extend the work of \cite{scemim} to design a dynamic dual-branch MI weighting mechanism.

{\bf Dynamic dual-branch MI weighting mechanism.} Given the hidden features\textsuperscript{\ref{ft:2}} $\mathbf{Z}_i^{\text{rec}}\in \mathbb{R}^{d}$ and $\mathbf{Z}_i^{\text{exp}}\in \mathbb{R}^{d}$ at time step $i$, which contain time-aware personalized information and fine-grained multi-interest information, we use two learnable weights $\mu_i^{\text{rec}}$ and $\mu_i^{\text{exp}}$ to adaptively learn MI contributions for the recommendation branch and the explanation branch, respectively, as follows:
\begin{small}
\begin{equation}
\mu_i^{\text{rec}} = \sigma(\mathrm{MLP}_{\text{rec}}(\mathbf{Z}_i^{\text{rec}})), \quad \mu_i^{\text{exp}} = \sigma(\mathrm{MLP}_{\text{exp}}(\mathbf{Z}_i^{\text{exp}})),
\end{equation}
\end{small}
where $\mathrm{MLP}_{\text{rec}}(\cdot):\mathbb{R}^{d}\rightarrow \mathbb{R}^{1}$ and $\mathrm{MLP}_{\text{exp}}(\cdot):\mathbb{R}^{d}\rightarrow \mathbb{R}^{1}$ , with sigmoid activation $\sigma(\cdot)$. We design the following objective function to make the model learn the semantic alignment information from the recommendation branch and the explanation branch as follows:
\begin{small}
\begin{equation}\label{eq:14}
\mathcal{J}_{\text{MI}}=\frac{1}{n}\sum\nolimits_{i=1}^n [ \mu_i^{\text{rec}} \cdot \mathcal{J}_{\text{MI}}^{\text{rec}} + \mu_i^{\text{exp}} \cdot \mathcal{J}_{\text{MI}}^{\text{exp}} ],
\end{equation}
\end{small}
where $\mathcal{J}_{\text{MI}}^{\text{rec}}=-\frac{1}{|\mathcal{V}|} \sum_{i=1}^{|\mathcal{V}|} \log \frac{\exp(\mathcal{S}^{\text{rec}}_i)}{\sum_{j=1}^{|\mathcal{V}|} \exp(\mathcal{S}^{\text{rec}}_j)}$ and $\mathcal{J}_{\text{MI}}^{\text{exp}}=-\frac{1}{|\mathcal{E}|} \sum_{i=1}^{|\mathcal{E}|} \log \frac{\exp(\mathcal{S}^{\text{exp}}_i)}{\sum_{j=1}^{|\mathcal{E}|} \exp(\mathcal{S}^{\text{exp}}_j)}$, which are proposed in \cite{scemim}. Here, $\mathcal{S}^{\text{rec}}_i=(\mathbf{M}_\mathcal{V})_i\Lambda^{\text{rec}}\mathbf{Z}_i^{\text{exp}}$ measures the similarity between $\mathbf{Z}_i^{\text{exp}}$ and the $i$-th item embedding $(\mathbf{M}_\mathcal{V})_i$, $\mathcal{S}^{\text{exp}}_i=(\mathbf{M}_\mathcal{E})_i\Lambda^{\text{exp}}\mathbf{Z}_i^{\text{rec}}$ measures the similarity between $\mathbf{Z}_i^{\text{rec}}$ and the $i$-th explanation embedding $(\mathbf{M}_\mathcal{E})_i$. These similarities are calculated by the two learnable matrices $\Lambda^{\text{rec}}$ and $\Lambda^{\text{exp}}$. In this way, the term $\mathcal{J}_{\text{MI}}^{\text{rec}}$ encourages each $\mathbf{Z}_i^{\text{exp}}$ to be closer to its paired recommendation embedding i.e. $(\mathbf{M}_\mathcal{V})_i$ than to unpaired ones. Same for $\mathcal{J}_{\text{MI}}^{\text{exp}}$ and each $\mathbf{Z}_i^{\text{rec}}$. 

\begin{table*}[!t]
\small
\setlength{\tabcolsep}{3pt}
\begin{tabular}{l|llll|llll|llll}
\hline
\multicolumn{1}{c|}{\multirow{3}{*}{Model}} & \multicolumn{4}{c|}{AM-Electronics}                                                                        & \multicolumn{4}{c|}{AM-Movies}                                                                             & \multicolumn{4}{c}{Yelp}                                                                                  \\ \cline{2-13} 
\multicolumn{1}{c|}{}                       & \multicolumn{2}{c}{Recommendation}                  & \multicolumn{2}{c|}{Explanation}                     & \multicolumn{2}{c}{Recommendation}                  & \multicolumn{2}{c|}{Explanation}                     & \multicolumn{2}{c}{Recommendation}                  & \multicolumn{2}{c}{Explanation}                     \\ \cline{2-13} 
\multicolumn{1}{c|}{}                       & \multicolumn{1}{c}{R@10} & \multicolumn{1}{c}{N@10} & \multicolumn{1}{c}{R@10} & \multicolumn{1}{c|}{N@10} & \multicolumn{1}{c}{R@10} & \multicolumn{1}{c}{N@10} & \multicolumn{1}{c}{R@10} & \multicolumn{1}{c|}{N@10} & \multicolumn{1}{c}{R@10} & \multicolumn{1}{c}{N@10} & \multicolumn{1}{c}{R@10} & \multicolumn{1}{c}{N@10} \\ \hline
PITF                                        & 0.0286                   & 0.0139                   & 0.0880                   & 0.0547                    & 0.0235                   & 0.0101                   & 0.0984                   & 0.0679                    & 0.0204                   & 0.0094                   & 0.0333                   & 0.0185                   \\
BPER                                        & 0.0307                   & 0.0154                   & 0.0901                   & 0.0553                    & 0.0281                   & 0.0131                   & 0.1028                   & 0.0687                    & 0.0212                   & 0.0106                   & 0.0393                   & 0.0214                   \\
SCEMIM-GRU4Rec                              & 0.0290                   & 0.0152                   & 0.0948                   & 0.0629                    & 0.0432                   & 0.0216                   & 0.1151                   & 0.0771                    & 0.0452                   & 0.0228                   & 0.0417                   & 0.0233                   \\
SCEMIM-SASRec                               & 0.0369                   & 0.0195                   & 0.1011                   & 0.0681                    & 0.0541                   & 0.0257                   & 0.1229                   & 0.0809                    & 0.0592                   & 0.0364                   & 0.0419                   & 0.0231                   \\
SCEMIM-FMLPRec                              & 0.0388                   & 0.0204                   & 0.1032                   & 0.0702                    & 0.0558                   & 0.0261                   & 0.1192                   & 0.0822                    & 0.0639                   & {\ul 0.0403}             & 0.0408                   & 0.0249                   \\
SCEMIM-LRURec                               & 0.0327                   & 0.0164                   & 0.1087                   & 0.0727                    & 0.0549                   & 0.0269                   & 0.1171                   & 0.0830                    & {\ul 0.0650}             & 0.0397                   & 0.0422                   & 0.0230                   \\
SCEMIM-CL4SRec                              & 0.0367                   & 0.0183                   & 0.1090                   & 0.0713                    & 0.0515                   & 0.026                    & 0.1213                   & 0.0831                    & 0.0536                   & 0.0320                   & 0.0424                   & 0.0259                   \\
SCEMIM-DiffuRec                             & {\ul 0.0402}             & {\ul 0.0212}             & 0.1186                   & {\ul 0.0794}              & 0.0544                   & 0.0282                   & {\ul 0.1352}             & 0.0897                    & 0.0624                   & 0.0377                   & {\ul 0.0498}             & {\ul 0.0276}             \\ \hline
SCEMIM-ComiRec                              & 0.0391                   & 0.0185                   & {\ul 0.1229}             & 0.0725                    & 0.0509                   & 0.0262                   & 0.1247                   & 0.0817                    & 0.0583                   & 0.0358                   & 0.0449                   & 0.0273                   \\
SCEMIM-Re4                                  & 0.0390                   & 0.0194                   & 0.1170                   & 0.0784                    & 0.0487                   & 0.0271                   & 0.1207                   & 0.0815                    & 0.0608                   & 0.0354                   & 0.0455                   & 0.0258                   \\
SCEMIM-TEDDY                                & 0.0385                   & 0.0206                   & 0.1173                   & 0.0735                    & 0.0514                   & 0.0275                   & 0.1201                   & 0.0854                    & 0.0599                   & 0.0367                   & 0.0432                   & 0.0260                   \\ \hline
SCEMIM-TLSAN                                & 0.0383                   & 0.0154                   & 0.0957                   & 0.0574                    & 0.0457                   & 0.0244                   & 0.1012                   & 0.0707                    & 0.0492                   & 0.0271                   & 0.0375                   & 0.0215                   \\
SCEMIM-MOJITO                               & 0.0399                   & 0.0201                   & 0.1156                   & 0.0690                    & 0.0508                   & 0.0280                   & 0.1245                   & 0.0840                    & 0.0572                   & 0.0354                   & 0.0424                   & 0.0262                   \\
SCEMIM-UniRec                               & 0.0401                   & 0.0211                   & 0.1201                   & 0.0790                    & {\ul 0.0560}             & {\ul 0.0301}             & 0.1339                   & {\ul 0.0924}              & 0.0597                   & 0.0386                   & 0.0473                   & 0.0270                   \\ \hline
\textbf{TME-PSR}                            & \textbf{0.0550}          & \textbf{0.0291}          & \textbf{0.1510}          & \textbf{0.0979}           & \textbf{0.0620}          & \textbf{0.0339}          & \textbf{0.1359}          & \textbf{0.0931}           & \textbf{0.0740}          & \textbf{0.0455}          & \textbf{0.0533}          & \textbf{0.0309}          \\
Improvement                                 & 36.82\%                  & 37.26\%                  & 22.86\%                  & 23.30\%                   & 10.71\%                  & 12.62\%                  & 0.52\%                   & 0.76\%                    & 13.87\%                  & 12.90\%                  & 7.03\%                   & 11.96\%                  \\ \hline
\end{tabular}
\caption{Recommendation and explanation performance, best results are marked in bold, second best results underlined. The improvement row is calculated as the percentage increase of the best result relative to the second best result.}
\label{tab:overall}
\end{table*}

\subsection{Optimization}

{\bf Prediction layer.} Given the hidden features $\mathbf{Z}^{\text{rec}}_i\in \mathbb{R}^{d}$ from \cref{eq:11} and $\mathbf{Z}^{\text{exp}}_i\in \mathbb{R}^{d}$ from \cref{eq:12} at time step $i$, the prediction layer contains two functions $\verb|Pred|(\mathbf{Z}^{\text{rec}}_i):\mathbb{R}^{d}\rightarrow \mathbb{R}^{|\mathcal{V}|}$ and $\verb|Pred|(\mathbf{Z}^{\text{exp}}_i):\mathbb{R}^{d}\rightarrow \mathbb{R}^{|\mathcal{E}|}$. The function $\verb|Pred|(\mathbf{Z}^{\text{rec}}_i)$ produces the probabilities of the user $u$ selecting the next item; the function $\verb|Pred|(\mathbf{Z}^{\text{exp}}_i)$ produces the degree of alignment between the item and all explanations:
\begin{small}
\begin{equation}
\verb|Pred|(\mathbf{Z}^{\text{rec}}_i)=\mathbf{M}_\mathcal{V}\mathbf{Z}^{\text{rec}}_i+\mathbf{b}^{\text{rec}}, \verb|Pred|(\mathbf{Z}^{\text{exp}}_i)=\mathbf{M}_\mathcal{E}\mathbf{Z}^{\text{exp}}_i+\mathbf{b}^{\text{exp}},
\end{equation}

\end{small}
where $\mathbf{b}^{\text{rec}}\in \mathbb{R}^{|\mathcal{V}|}$ and $\mathbf{b}^{\text{exp}}\in \mathbb{R}^{|\mathcal{E}|}$ are bias terms. Thus, the complete optimization objective is given as follows:
\begin{small}
\begin{equation}
\mathcal{L} = \mathcal{L}_{\text{rec}} + \mathcal{L}_{\text{exp}} + \mathcal{J}_{\text{MI}}.
\end{equation}
\end{small}

The recommendation objective $\mathcal{L}_{\text{rec}}$ and the explanation objective $\mathcal{L}_{\text{exp}}$ are computed via cross-entropy loss based on the prediction scores $\verb|Pred|(\mathbf{Z}^{\text{rec}})$ and $\verb|Pred|(\mathbf{Z}^{\text{exp}})$.\footnote{The formulations of $\mathcal{L}_{\text{rec}}$ and $\mathcal{L}_{\text{exp}}$ are provided in App. B.1.}

By optimizing the final objective $\mathcal{L}$, the proposed TME-PSR model achieves time-aware personalization, multi-interest personalization and explanation personalization for sequential recommendation, thereby improving both recommendation and explanation performance.

\begin{table}[t]
\fontsize{8pt}{10pt}\selectfont
\setlength{\tabcolsep}{2pt}
\begin{tabular}{ccc|llll}
\hline
\multicolumn{3}{c|}{\multirow{2}{*}{Strategy}}                                                         & \multicolumn{4}{c}{AM-Electronics}                                    \\ \cline{4-7} 
\multicolumn{3}{c|}{}                                                                                  & \multicolumn{2}{c}{Rec}           & \multicolumn{2}{c}{Exp}           \\ \hline
\multicolumn{1}{l}{Time-aware} & \multicolumn{1}{l}{Multi-interest} & \multicolumn{1}{l|}{Explanation} & R@10            & N@10            & R@10            & N@10            \\ \hline
                               &                                    &                                  & 0.0327          & 0.0164          & 0.1087          & 0.0727          \\ \cline{1-3}
\Checkmark                     &                                    &                                  & 0.0402          & 0.0197          & 0.1371          & 0.0912          \\
                               & \Checkmark                         &                                  & 0.0361          & 0.0184          & 0.1312          & 0.0849          \\
                               &                                    & \Checkmark                       & 0.0324          & 0.0160          & 0.1252          & 0.0844          \\ \cline{1-3}
\Checkmark                     & \Checkmark                         &                                  & {\ul 0.0498}    & {\ul 0.0247}    & {\ul 0.1455}    & {\ul 0.0938}    \\
\Checkmark                     &                                    & \Checkmark                       & 0.0450          & 0.0240          & 0.1423          & 0.0920          \\
                               & \Checkmark                         & \Checkmark                       & 0.0401          & 0.0234          & 0.1378          & 0.0915          \\ \cline{1-3}
\Checkmark                     & \Checkmark                         & \Checkmark                       & \textbf{0.0550} & \textbf{0.0291} & \textbf{0.1510} & \textbf{0.0979} \\ \hline
\end{tabular}
\caption{Ablation of key strategies the AM-Electronics dataset, best results are marked in bold,
second best results underlined.}
\label{tab:abl}
\end{table}

\section{EXPERIMENTS}

{\bf Models and datasets.} We compare the proposed TME-PSR model with 4 types of
 SR approaches: (1) {\it General models}: GRU4Rec \cite{gru4rec}, SASRec \cite{sasrec}, FMLPRec \cite{fmlprec}, CL4SRec \cite{cl4rec}, LRURec \cite{lrurec}, and DiffuRec \cite{diffurec}. (2) {\it Models with temporal modeling}:  TLSAN \cite{tlsan}, MOJITO \cite{mojito} and UniRec \cite{unirec}. (3) {\it Models with multi-interest modeling}: ComiRec \cite{comirec}, Re4 \cite{re4} and TEDDY \cite{teddy}. (4) {\it Explainable models}: PITF \cite{pitf} and BPER \cite{bper}. Since the first three types of models lack explainability, we insert these 12 models as backbones into the explainable SR framework SCEMIM \cite{scemim} (The experiments described in SCEMIM have demonstrated that it can enable the model to acquire explainability while maintaining or even improving recommendation performance), thus obtaining 12 explainable SR models, denoted by SCEMIM-X (where X represents each model).
 
 We conduct experiments on three widely used datasets for explainable SR, including the Amazon Movies dataset, the Amazon Electronics dataset, and the Yelp \cite{dataset} dataset. In these datasets, each user-item interaction is associated with a timestamp, an item ID, and an explanation ID. The explanations are extracted from user reviews and clustered into representative phrases through semantic hashing techniques \cite{dataset}. We follow the previous sequential recommendation work \cite{sasrec,scemim} to divide each user’s interaction history  chronologically into three parts: the most recent interaction for testing, the second most recent for validation, and the remaining for training.\footnote{\label{ft:3}The statistics of the preprocessed datasets, the hyperparameter analysis and more implementation details are provided in App. B.} We evaluate both recommendation and explanation performance using two standard ranking metrics: Recall@K and NDCG@K, where K $=10$. We refer to Recall@K as R@K and NDCG@K as N@K. We set hyperparameters $\alpha=0.9$, $\beta=0.1$, $d=50$ and $H=2$.\textsuperscript{\ref{ft:3}} 

{\bf Recommendation and explanation performance comparison.} Tab.~\ref{tab:overall} shows the overall performance of all methods in recommendation and explanation across three datasets. As shown, our method consistently outperforms all baseline models across all evaluation metrics for both the recommendation task from 10.71\% to 37.26\% and the explanation tasks from 0.52\% to 25.73\%. These results verify the wide applicability and stability of the proposed model. A comparison with the baseline models without SCEMIM are provided in App. C. Results show that our method still provides a significant improvement.

{\bf Ablation Study.} 
To further investigate the effectiveness of the three proposed personalized strategies, we evaluate every non-empty combination of the three personalization strategies on the AM-Electronics dataset.\footnote{The ablation study on all three datasets are provided in App. D.} The main findings are as follows: (1) All three strategies are individually beneficial. Activating any single component (rows 2–4) already lifts both R@10 and N@10 above the backbone with no personalization (row 1). (2) Pairwise combinations show clear synergy. For example, combining time-aware personalization and multi-interest personalization (row 5) outperforms any single component. (3) The full model is consistently the best. Enabling all three strategies (bottom row) attains the top performance on every metric. This confirms that the three types of personalization have significant mutual synergy. In summary, the complete TME-PSR configuration reliably delivers the strongest recommendation and explanation quality across all settings.

\begin{figure}[t]
    \centering
    \includegraphics[width=0.8\linewidth]{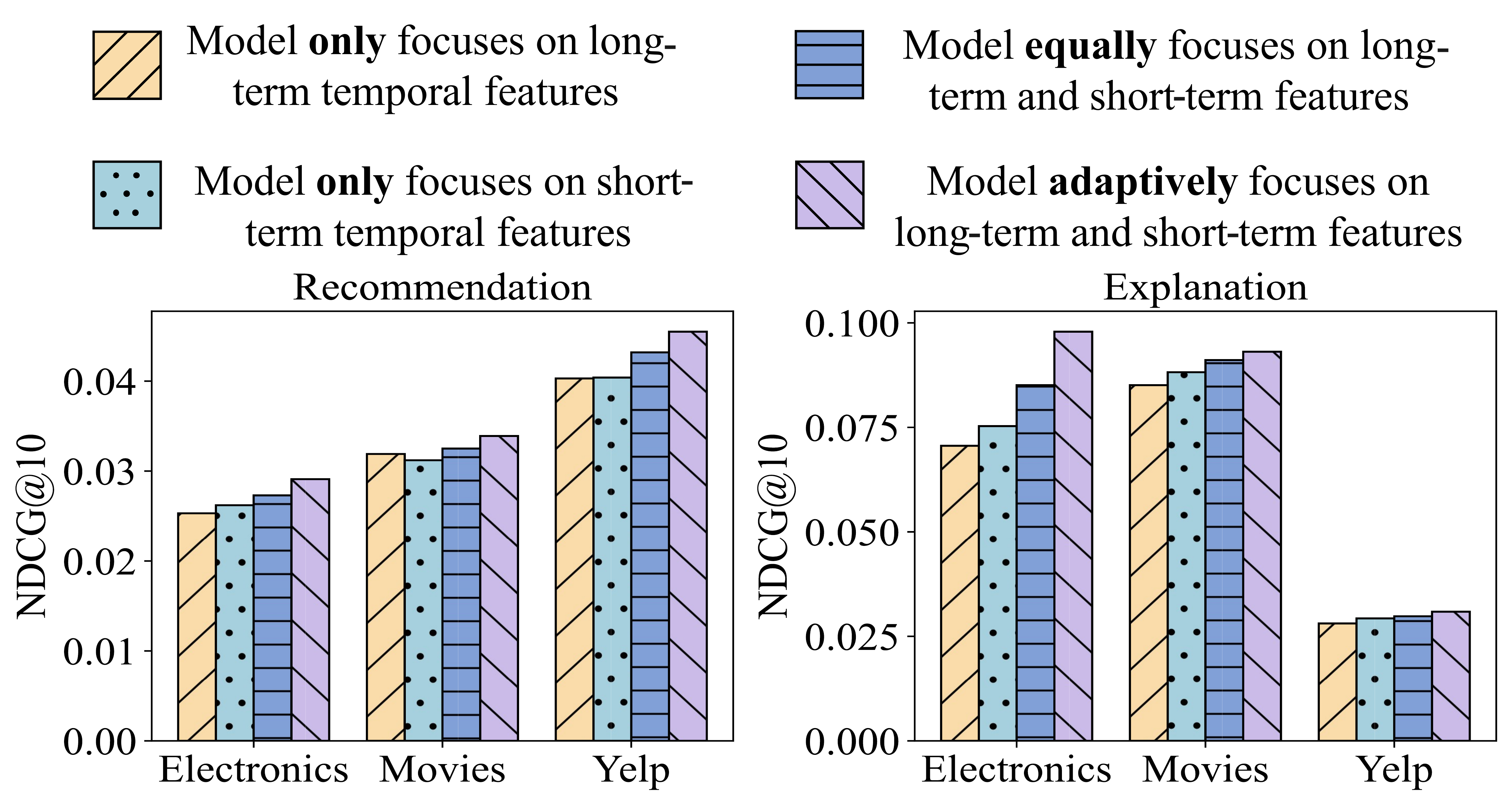}
    \caption{Comparison of four gating strategies. Results show that the model that adaptively focuses on both long-term and short-term features consistently achieves the best performance, which proves the effectiveness of the proposed method.}
    \label{fig:gamma_abl}
\end{figure}

{\bf Detailed analysis of the effectiveness of the time-aware personalization.} In \cref{eq:3} we use a gating parameter $\gamma^{\text{rec}}$ to control the contribution of short-term and long-term temporal features to the recommendation time-aware embedding $\tilde{\mathbf{E}}^{\text{rec}}$. A larger $\gamma^{\text{rec}}$ means a greater contribution of short-term temporal features to $\tilde{\mathbf{E}}^{\text{rec}}$, while a smaller $\gamma^{\text{rec}}$ emphasizes long-term temporal features in $\tilde{\mathbf{E}}^{\text{rec}}$. Same for the gating parameter $\gamma^{\text{exp}}$ and the explanation time-aware embedding $ \tilde{\mathbf{E}}^{\text{exp}}$ in \cref{eq:4}. Such a dual-view gated time encoder helps us realize the time-aware personalization.

To further validate the effectiveness of our dual-view gated time encoder, we compare the following four baselines: (1) We set $\tilde{\mathbf{E}}^{\text{rec}}=\tilde{\mathbf{E}}^{\text{exp}}=\mathbf{H}^{\text{abs}}$ to obtain a model that only focuses on long-term temporal features. (2) We set $\tilde{\mathbf{E}}^{\text{rec}}=\tilde{\mathbf{E}}^{\text{exp}}=\mathbf{H}^{\text{adj}}$ to obtain a model that only focuses on short-term temporal features. (3) We set $\tilde{\mathbf{E}}^{\text{rec}}=\tilde{\mathbf{E}}^{\text{exp}}=0.5\times(\mathbf{H}^{\text{adj}} + \mathbf{H}^{\text{abs}})$ to obtain a model that focuses on short-term and long-term temporal features equally. (4) A model uses the proposed dual-view gated time encoder, which focuses on short-term and long-term temporal features adaptively as \cref{eq:3} and \cref{eq:4} show. As shown in \cref{fig:gamma_abl}, models that focus on both short-term and long-term temporal patterns outperform those that only focus on a single pattern, and dynamically balancing the two patterns further improves performance, validating the effectiveness of the proposed dual-view gated time encoder. We also found a linear relationship between the values of the two gating parameters and the average time interval of each user sequence, please see App. E for details.

\begin{figure}[!t]
    \centering
    \includegraphics[width=0.8\linewidth]{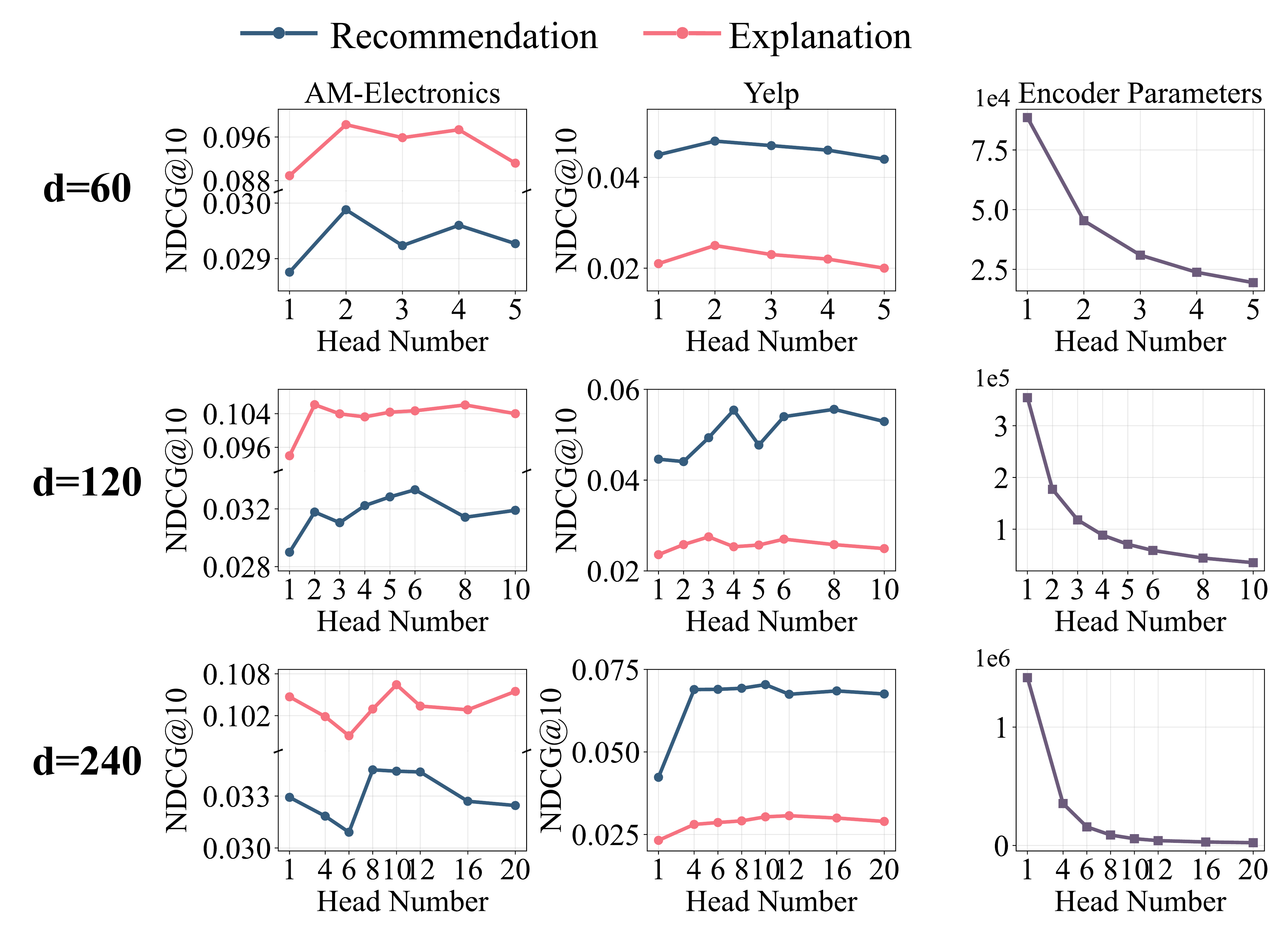}
    \caption{Left two columns show recommendation and explanation performance with different head numbers $H$ under different embedding sizes $d$. Results show that the optimal head number increases with the increase of $d$. The third column shows the relationship between $H$ and the encoder parameters. As $H$ increases, the number of parameters decreases inversely.}
    \label{fig:headnum}
\end{figure}

{\bf Detailed analysis of the effectiveness of the multi-interest personalization.} 
To further explore the effect of the head number $H$ of the proposed multihead LRU architecture, we evaluate the impact of different head numbers $H$ under varying embedding sizes $d \in\{ 60, 120, 240\}$, ensuring that the head number $H$ is a divisor of $d$. As \cref{fig:headnum} shows, the optimal head number $H^*$ consistently increases with the performance on both the recommendation branch and the explanation branch: $H^*\approx2,6,10$ for $d=60,120,240$, and as the embedding size $d$ increases, the overall performance improves. These findings indicate that higher dimensions of the representation space provide richer capabilities for interest disentanglement, and higher dimensions require more fine-grained disentanglement. When $H$ exceeds $H^*$, performance begins to degrade, likely due to each head being allocated too narrow a subspace, limiting its expressiveness. We also explored the relationship between the number of encoder parameters and the head number $H$. As the third column of \cref{fig:headnum} shows, when $d$ remains fixed, the model parameters gradually decrease as $H$ increases, which follows an inverse proportional relationship $\frac{d^2}{H}$ and achieving stronger disentanglement with fewer parameters as subspaces get finer.

\begin{figure}[t]
    \centering
    \includegraphics[width=0.9\linewidth]{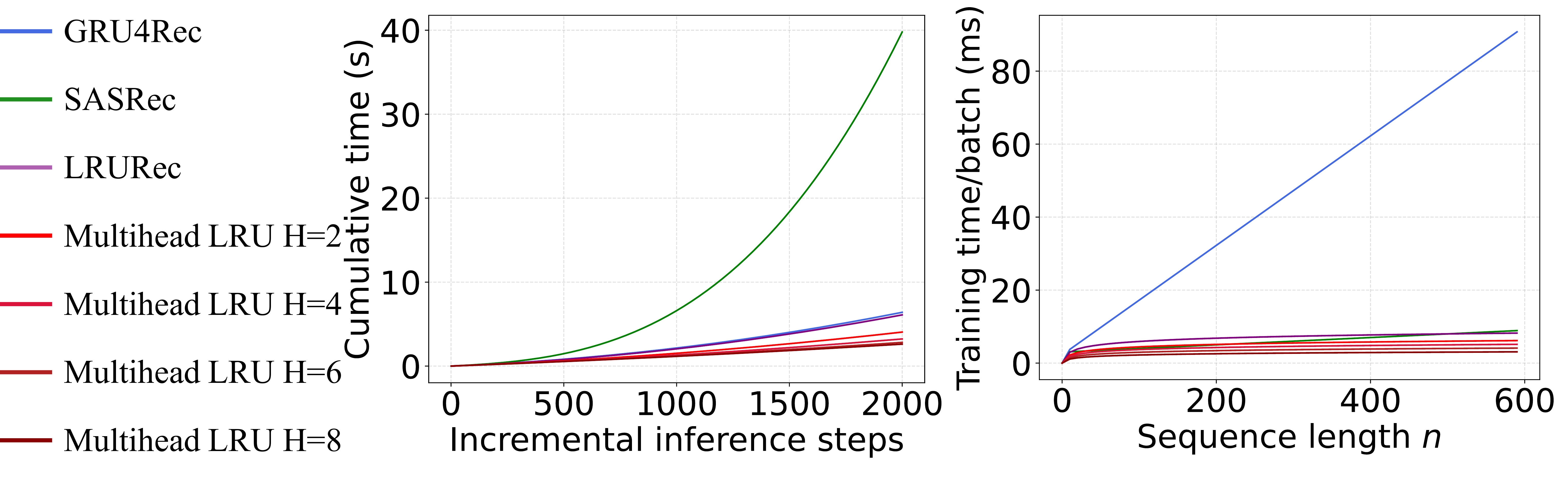}
    \caption{Inference and training efficiency of GRU4Rec, SASRec, LRURec and the proposed multihead LRU. Results show that the proposed method outperforms the other three methods, and its efficiency increases with the head number $H$.}
    \label{fig:efficiency}
\end{figure}

We further explored incremental inference and training efficiency across GRU4Rec \cite{gru4rec}, SASRec \cite{sasrec}, LRURec \cite{lrurec}, and the proposed multihead LRU ($H=2,4,6,8$) on embedding size $d=240$, batch\_size$=256$. We perform only one batch of inference at each step and continuously expand the input length after each prediction, and test the training time required for a batch. As \cref{fig:efficiency} shows, the proposed method outperforms the compared three methods in terms of incremental inference and training efficiency, and the performance improvement is more significant with a larger number of heads. 

In summary, as $d$ increases, accordingly increases the number of heads $H$ simultaneously deliver: (1) higher accuracy (2) reduced parameters (3) faster encoding. These results provide converging evidence that our approach achieves fine-grained sub-interest disentanglement while remaining lightweight and efficient.

\begin{figure}[!t]
    \centering
    \includegraphics[width=1\linewidth]{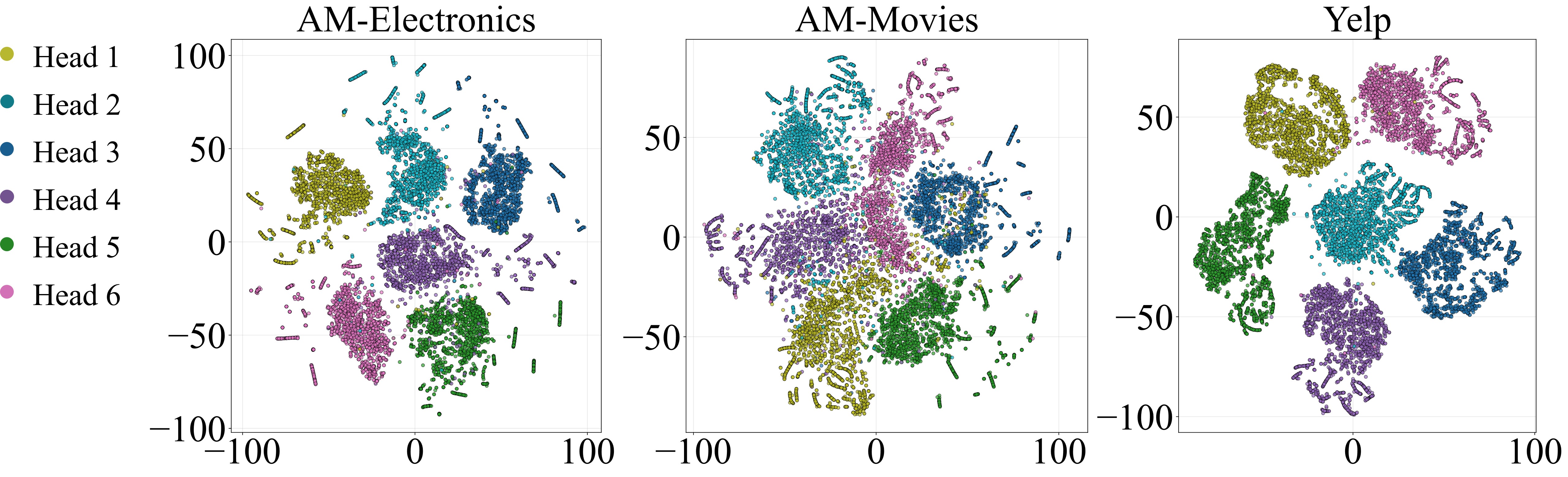}
    \caption{Visualization of multihead LRU output representation at $d=120$, $H=6$ with t-SNE.}
    \label{fig:head_clu}
\end{figure}

To further explore whether the proposed multihead LRU architecture truly capture distinct and fine-grained sub-interests, we performed a t-SNE \cite{tsne} analysis on the output representations of each head under the conditions of $d=120$ and $H=6$. As \cref{fig:head_clu} shows, the six heads form well-separated clusters; each head covers a complementary region in the embedding space.

\begin{figure}[!t]
    \centering
    \includegraphics[width=0.8\linewidth]{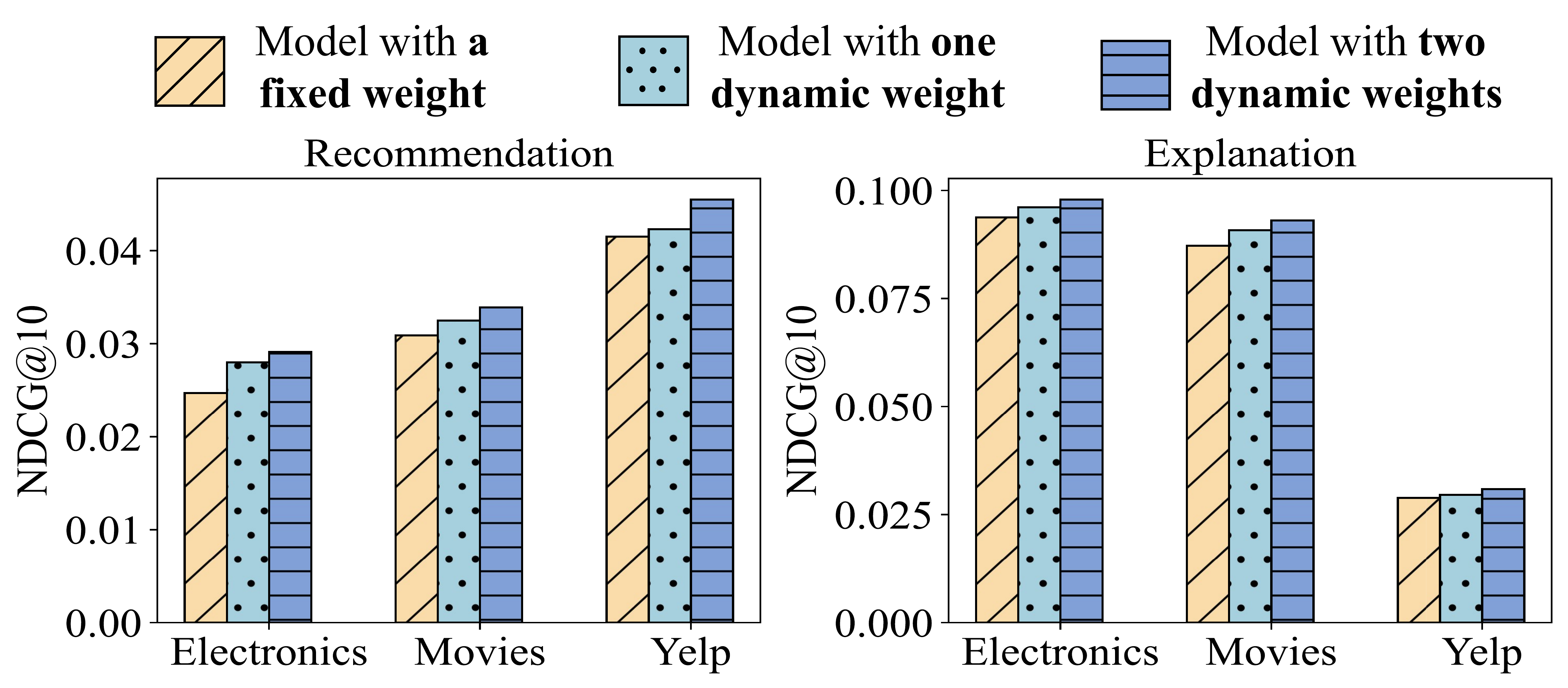}
    \caption{Comparison of three MI weighting strategies. Results show that the model with two dynamic weights consistently achieves the best performance, which proves the effectiveness of the proposed dynamic dual-branch MI weighting mechanism.}
    \label{fig:mu_abl}
\end{figure}

{\bf Detailed analysis of the effectiveness of the explanation personalization.} 
In \cref{eq:14}, we use two learnable weights $\mu^{\text{rec}}$ and $\mu^{\text{exp}}$ to dynamically control the contribution of the recommendation MI loss $\mathcal{J}_{\text{MI}}^{\text{rec}}$ and the explanation MI loss $\mathcal{J}_{\text{MI}}^{\text{exp}}$ to the total MI loss $\mathcal{J}_{\text{MI}}$. To further validate the effectiveness of such a dynamic dual-branch MI weighting mechanism, we compare the following three baselines: (1) We set $\mathcal{J}_{\text{MI}}=0.001\times(\mathcal{J}_{\text{MI}}^{\text{rec}} + \mathcal{J}_{\text{MI}}^{\text{exp}})$ to obtain a model that uses a fixed weight for $\mathcal{J}_{\text{MI}}^{\text{rec}}$ and $\mathcal{J}_{\text{MI}}^{\text{exp}}$. (2) We set $\mathcal{J}_{\text{MI}}=\mu\cdot(\mathcal{J}_{\text{MI}}^{\text{rec}} + \mathcal{J}_{\text{MI}}^{\text{exp}})$ with $\mu= \sigma(\mathrm{MLP}([\mathbf{Z}_i^{\text{rec}},\mathbf{Z}_i^{\text{exp}}]))$ to obtain a model that uses one learnable weight for $\mathcal{J}_{\text{MI}}^{\text{rec}}$ and $\mathcal{J}_{\text{MI}}^{\text{exp}}$. (3) A model using the proposed dynamic MI weighting mechanism, which applies two dynamic weights $\mu^{\text{rec}}$ and $\mu^{\text{exp}}$ to $\mathcal{J}_{\text{MI}}^{\text{rec}}$ and $\mathcal{J}_{\text{MI}}^{\text{exp}}$, respectively, as \cref{eq:14} shows. As \cref{fig:mu_abl} shows, the model with learnable weights outperform those with a fixed weight, and dynamically learning two task-specific MI weights further improves performance, validating the effectiveness of the proposed dynamic dual-branch MI weighting mechanism.

\begin{figure}[!t]
    \centering
    \includegraphics[width=0.8\linewidth]{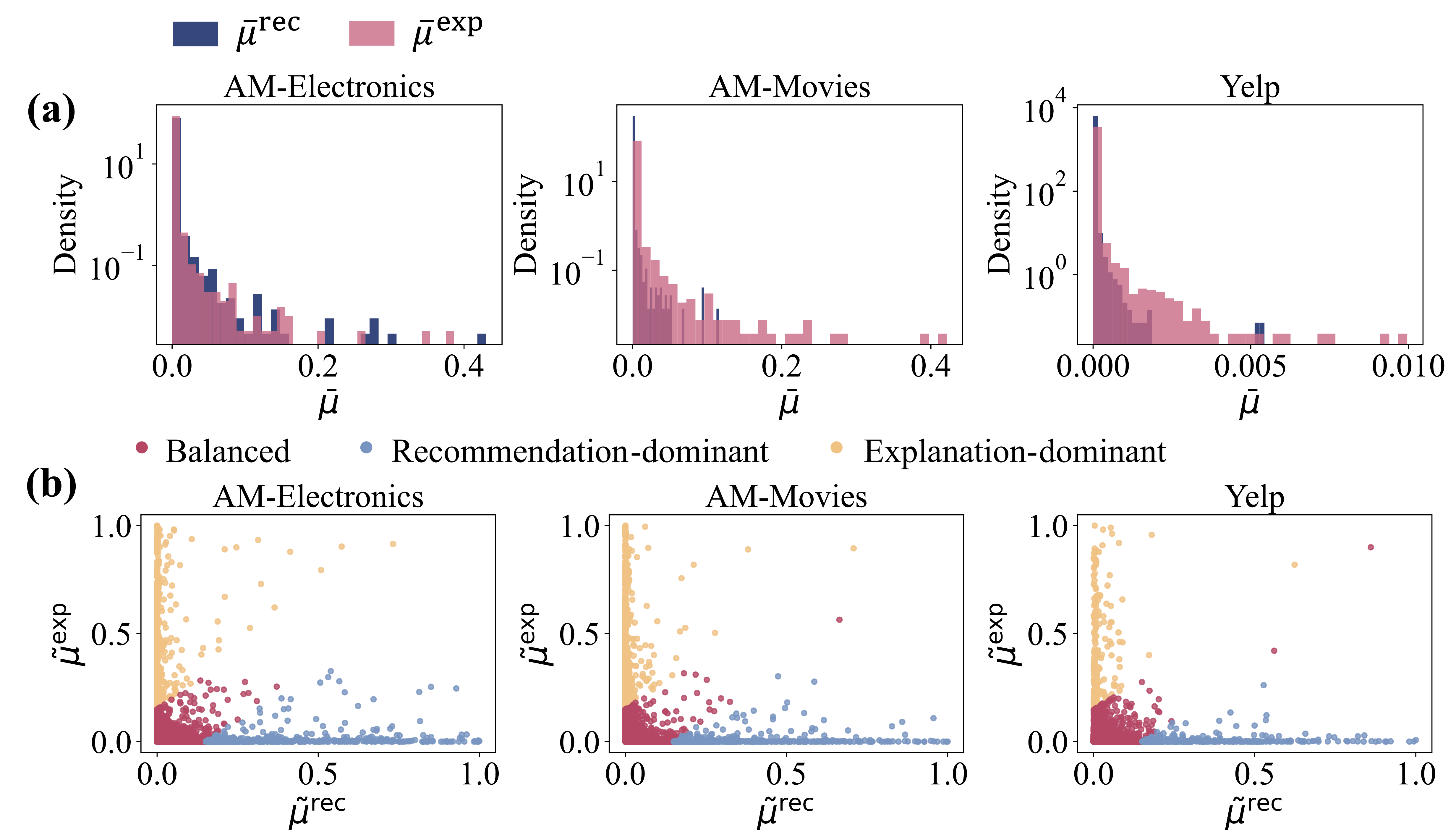}
    \caption{(a) Visualization of distributions of all interaction sequences average dynamic MI weights $\bar{\mu}^{\text{rec}}$ and $\bar{\mu}^{\text{exp}}$. 
    (b) K-means clustering based on $(\tilde{\mu}^{\text{rec}},\tilde{\mu}^{\text{exp}})$. 
    }
    \label{fig:mu_dis}
\end{figure}

Let $\mu^{\text{rec}}_{(u,i)}$ represent the MI weight of the recommendation task of the $i$-th interaction of user sequence $S_u$, and $\mu^{\text{exp}}_{(u,i)}$ represent the MI weight of the explanation task. We compute the average weight $\bar{\mu}^{\text{rec}}_u=\sum^n_{i=1}\mu^{\text{rec}}_{(u,i)}/n$ and $\bar{\mu}^{\text{exp}}_u=\sum^n_{i=1}\mu^{\text{exp}}_{(u,i)}/n$. \cref{fig:mu_dis}(a) reports the distributions of the averaged weights $\bar{\mu}^{\text{rec}}_u$ and $\bar{\mu}^{\text{exp}}_u$ of all interaction sequences. Results show that all distributions are highly left-skewed and concentrated near zero, indicating that most sequences are assigned low semantic alignment strengths between recommendations and explanations in both the recommendation task and the explanation task.\footnote{Specific examples of the two types of user alignment requirements discussed in \cref{fig:mu_dis}(a) and the three types of users in \cref{fig:mu_dis}(b) are provided in App. G.\label{ft:appf}}

To further investigate user-level differences in the degree of semantic alignment strength, we apply an enhanced K-means clustering \cite{kmeans} on the following 2D points set, $(\tilde{\mu}^{\text{rec}}_u,\tilde{\mu}^{\text{exp}}_u)$, where $\tilde{\mu}^{\text{rec}}_u$ and $\tilde{\mu}^{\text{exp}}_u$ are normalized values of $\bar{\mu}^{\text{rec}}_u$ and $\bar{\mu}^{\text{exp}}_u$. Results show that user-level $\tilde{\mu}^{\text{rec}}_u$ and $\tilde{\mu}^{\text{exp}}_u$ exhibit three well-separated clusters shown in \cref{fig:mu_dis}(b): (1) A balanced group near the diagonal, this type of users has similar semantic alignment requirements in the recommendation task and the explanation task. (2) A recommendation-dominant group, this type of users has higher semantic alignment requirements in the recommendation task. (3) An explanation-dominant group, this type of users has higher semantic alignment requirements in the explanation task.\footref{ft:appf} This proves that even for the same user, the recommendation task and the explanation task have different requirements for semantic alignment, which confirms the necessity of personalization.

\section{CONCLUSION}

This paper proposes TME-PSR, a model for sequential recommendation that simultaneously addresses three types of personalization: time-aware, multi-interest, and explanation personalization. Our approach advances the field by demonstrating that personalization should be considered jointly across multiple dimensions. Extensive experiments demonstrate that our model consistently improves recommendation accuracy, explanation quality and efficiency across multiple datasets. For future work, we plan to extend personalization to cross-domain or multi-modal recommendation settings, while maintaining explainability, represents an exciting opportunity to generalize the proposed model further.

\newpage
{
    \bibliographystyle{named}
    \bibliography{ijcai26}
}

\newpage
\appendix

\section{Additional Related Work}\label{app:A}
Sequential recommendation predicts a user's next interaction by modeling their historical behaviors. Early models used Markov chains \cite{markov} to model sequential data. Early studies adopted CNN \cite{cnnrec,caser,cnnrec2} and RNN methods \cite{rnnrec,gru4rec,gru4rec+,krnnp} to capture complex patterns. More recently, Transformer-based models \cite{sasrec,bert4rec,transformers4rec} use self-attention for long-range dependencies with better parallelism. Some studies used contrastive learning \cite{cl4rec,duorec,icsrec} to improve recommendation performance. Diffusion brings a generative view to SR by treating the target item as a distribution and denoising it from noise conditioned on a user’s history, which helps capture uncertainty and multi-aspect signals,  representative methods \cite{diffurec,dreamrec,diff4rec,diffuasr} show competitive accuracy with a practical denoising pipeline. GNN-based SR \cite{autoseqrec,mssg,dgsr} constructs session/transition graphs to encode higher-order item transitions and local dependencies. All-MLP designs \cite{fmlprec,trimlp,mlm4rec} replace self-attention with learnable filter blocks for high throughput and low latency. Multi-modal SR pretrains on text/images (etc.) and transfers to downstream SR: \cite{missrec} adopts a pretrain-and-transfer encoder–decoder with contrastive objectives, while \cite{unisrec} learns ID-agnostic universal sequence/item representations from descriptions for cross-domain transfer. Generative retrieval replaces ANN with sequence models that decode semantic IDs of items; \cite{tiger,mbgen,eager} demonstrates strong performance and unifies retrieval/ranking under a single autoregressive model. LLMs unify diverse rec tasks via text-to-text formulations \cite{p5,genrec,pod}, or use LLMs for learning semantic embeddings \cite{llmemb,llmesr,e4srec}. Handling ultra-long histories (thousands of events) motivates scalable encoders and training/serving optimizations; recent industry-scale methods \cite{lrurec,ubr4ctr,sam,mimn} design long-sequence-optimized Transformers and serving pipelines, and surveys position ultra-long SR as a key frontier.

\section{Implementation Details}

All experiments are conducted on a server equipped with 4×NVIDIA RTX 4090 GPUs (24GB each), an Intel Xeon Silver 4314 CPU @ 2.40GHz, and 256GB RAM. The software environment includes Ubuntu 18.04.4 LTS, Python 3.9.21, PyTorch 2.6.0+cu124, and CUDA 12.2.

\subsection{Complete formulations of the recommendation objective function and the explanation objective function. }
\begin{equation}
\mathcal{L}_{\text{rec}} = -\log \frac{\exp([\text{Pred}(\mathbf{Z}^{\text{rec}}_i)]_{y^{\text{rec}}_i})}{\sum_{v\in \mathcal{V}} \exp([\text{Pred}(\mathbf{Z}^{\text{rec}}_i)]_v)},
\end{equation}
\begin{equation}
\mathcal{L}_{\text{exp}} = -\log \frac{\exp([\text{Pred}(\mathbf{Z}^{\text{exp}}_i)]_{y^{\text{exp}}_i})}{\sum_{e\in \mathcal{E}} \exp([\text{Pred}(\mathbf{Z}^{\text{exp}}_i)]_e)},
\end{equation}
where $y^{\text{rec}}_i$ is the target label at time step $i$ in the recommendation branch; $y^{\text{exp}}_i$ is the target label at time step $i$ in the explanation branch.

\begin{table}[t]
\small
\centering
\setlength{\tabcolsep}{3pt}
\begin{tabular}{llll}
\hline
Dataset                   & Movies & Electronics & Yelp      \\ \hline
\# Users                  & 22,248        & 18,129             & 90,742    \\
\# Items                  & 15,458        & 10,435             & 46,372    \\
\# Explanations           & 31,451        & 23,541             & 119,874   \\
\# Actions                & 305,801       & 144,772            & 1,204,515 \\
Avg. actions/users        & 13.75         & 7.95               & 13.27     \\
Avg. actions/items        & 19.78         & 13.87              & 25.98     \\
Avg. actions/expls & 9.72          & 6.15               & 10.05     \\
User-item density         & 0.089\%       & 0.076\%            & 0.029\%   \\
User-expl density         & 0.063\%       & 0.034\%            & 0.022\%   \\
User-item-expl density    & 2.83e-6\%     & 3.23e-6\%          & 2.39e-7\% \\ \hline
\end{tabular}
\caption{The statistics of the preprocessed datasets.}
\label{tab:datasets}
\end{table}

\subsection{Baselines}

We compare our method with several representative sequential recommendation baselines:\\
$\bullet$ PITF \cite{pitf}: A pairwise-interaction tensor-factorization model using triadic (user, item, tag) embeddings to support explainable ranking.\\
$\bullet$ BPER \cite{bper}: A Bayesian-personalized explanation-ranking model using joint BPR optimization to align item recommendations with textual rationales.\\
$\bullet$ GRU4Rec \cite{gru4rec}: A gated-recurrent-unit \cite{gru} model using pairwise ranking losses to learn within-session user–item transitions.\\ 
$\bullet$ SASRec \cite{sasrec}: A transformer-based sequential model leveraging self-attention to capture user-item transition patterns.\\
$\bullet$ FMLPRec \cite{fmlprec}: A filter-based all-MLP model using learnable Fourier kernels to model sequential signals with low latency.\\
$\bullet$ LRURec \cite{lrurec}: A linear-recurrence model using closed-form state transitions to approximate RNN behavior while enabling parallel computation. \\
$\bullet$ CL4SRec \cite{cl4rec}: A contrastive-learning SR model that augments sequences and jointly optimizes next-item prediction with self-supervised contrastive signals to learn stronger user representations.\\
$\bullet$ DiffuRec \cite{diffurec}: A diffusion-based generative SR model that represents target items as distributions and denoises them from Gaussian noise conditioned on the user’s history, improving recommendation under preference uncertainty.\\
$\bullet$ ComiRec \cite{comirec}: A controllable multi-interest framework that extracts multiple intent vectors (via capsule/attention variants) and matches candidates to the most relevant interest for each user.\\
$\bullet$ Re4 \cite{re4}: A multi-interest framework using re-contrast, re-attend and re-construct regularisers to maintain distinct intent vectors.\\
$\bullet$ TEDDY \cite{teddy}: A disentangled SR method that separates interest trend and interest diversity (via adaptive masking and dedicated objectives) to improve robustness and personalization.\\
$\bullet$ TLSAN \cite{tlsan}: A dual-stream attention model using time-position embeddings for temporal drift.\\
$\bullet$ MOJITO \cite{mojito}: A time-aware Transformer that mixes temporal contexts with Gaussian-mixture attention, enabling better next-item prediction under varying temporal conditions.\\
$\bullet$ UniRec \cite{unirec}: A uniformity–frequency enhanced SR model that exploits regularized time spacing and item-frequency cues via dual objectives to mitigate drift and sparsity.

In addition, SCEMIM \cite{scemim} is an explainable sequential recommendation framework that enhances sequential recommendation with mutual information maximization between recommendation and explanation representations. 

\cref{tab:datasets} shows the statistics of the datasets.

\begin{table}[t]
\small
\setlength{\tabcolsep}{2pt}
\begin{tabular}{l|ll|ll|ll}
\hline
\multicolumn{1}{c|}{\multirow{3}{*}{Model}} & \multicolumn{2}{c|}{AM-Electronics}                  & \multicolumn{2}{c|}{AM-Movies}                       & \multicolumn{2}{c}{Yelp}                            \\ \cline{2-7} 
\multicolumn{1}{c|}{}                       & \multicolumn{2}{c|}{Recommendation}                  & \multicolumn{2}{c|}{Recommendation}                  & \multicolumn{2}{c}{Recommendation}                  \\ \cline{2-7} 
\multicolumn{1}{c|}{}                       & \multicolumn{1}{c}{R@10} & \multicolumn{1}{c|}{N@10} & \multicolumn{1}{c}{R@10} & \multicolumn{1}{c|}{N@10} & \multicolumn{1}{c}{R@10} & \multicolumn{1}{c}{N@10} \\ \hline
GRU4Rec                                     & 0.0259                   & 0.0134                    & 0.0431                   & 0.0208                    & 0.0451                   & 0.0228                   \\
SASRec                                      & 0.0342                   & 0.0182                    & 0.0536                   & 0.0256                    & 0.0604                   & 0.0364                   \\
FMLPRec                                     & 0.0357                   & 0.0186                    & 0.0558                   & 0.0253                    & 0.0620                   & 0.0403                   \\
LRURec                                      & 0.0345                   & 0.0159                    & {\ul 0.0560}             & 0.0259                    & 0.0625                   & 0.0397                   \\
CL4SRec                                     & 0.0371                   & 0.0194                    & 0.0523                   & 0.0262                    & 0.0510                   & 0.0318                   \\
DiffuRec                                    & 0.0389                   & 0.0201                    & 0.0548                   & 0.0273                    & 0.0599                   & 0.0325                   \\ \hline
ComiRec                                     & 0.0395                   & 0.0198                    & 0.0515                   & 0.0271                    & 0.0562                   & 0.0333                   \\
Re4                                         & 0.0355                   & 0.0185                    & 0.0540                   & 0.0265                    & {\ul 0.0643}             & 0.0400                   \\
TEDDY                                       & 0.0392                   & {\ul 0.0208}              & 0.0537                   & 0.0279                    & 0.0628                   & {\ul 0.0411}             \\ \hline
TLSAN                                       & 0.0372                   & 0.0191                    & 0.0464                   & 0.0245                    & 0.0571                   & 0.0323                   \\
MOJITO                                      & 0.0386                   & 0.0198                    & 0.0519                   & 0.0302                    & 0.0575                   & 0.0364                   \\
UniRec                                      & {\ul 0.0403}             & 0.0200                    & 0.0557                   & {\ul 0.0317}              & 0.0578                   & 0.0362                   \\ \hline
\textbf{TME-PSR}                            & \textbf{0.0550}          & \textbf{0.0291}           & \textbf{0.0620}          & \textbf{0.0339}           & \textbf{0.0740}          & \textbf{0.0455}          \\
Improvement                                 & 36.47\%                  & 39.90\%                   & 10.71\%                  & 6.94\%                    & 15.09\%                  & 10.71\%                  \\ \hline
\end{tabular}
\caption{overall with non-SCEMIM models}
\label{tab:overall_onlyrec}
\end{table}

\begin{table*}[!t]
\small
\setlength{\tabcolsep}{2pt}
\begin{tabular}{ccc|llllllllllll}
\hline
\multicolumn{3}{c|}{\multirow{2}{*}{Strategy}}                                                         & \multicolumn{4}{c}{AM-Electronics}                                    & \multicolumn{4}{c}{AM-Movies}                                         & \multicolumn{4}{c}{Yelp}                                              \\ \cline{4-15} 
\multicolumn{3}{c|}{}                                                                                  & \multicolumn{2}{c}{Rec}           & \multicolumn{2}{c}{Exp}           & \multicolumn{2}{c}{Rec}           & \multicolumn{2}{c}{Exp}           & \multicolumn{2}{c}{Rec}           & \multicolumn{2}{c}{Exp}           \\ \hline
\multicolumn{1}{l}{Time-aware} & \multicolumn{1}{l}{Multi-interest} & \multicolumn{1}{l|}{Explanation} & R@10            & N@10            & R@10            & N@10            & R@10            & N@10            & R@10            & N@10            & R@10            & N@10            & R@10            & N@10            \\ \hline
                               &                                    &                                  & 0.0327          & 0.0164          & 0.1087          & 0.0727          & 0.0549          & 0.0269          & 0.1171          & 0.0830          & 0.0650          & 0.0397          & 0.0422          & 0.0230          \\ \cline{1-3}
\Checkmark                     &                                    &                                  & 0.0402          & 0.0197          & 0.1371          & 0.0912          & 0.0563          & 0.0266          & 0.1287          & 0.0841          & 0.0672          & 0.0410          & 0.0454          & 0.0250          \\
                               & \Checkmark                         &                                  & 0.0361          & 0.0184          & 0.1312          & 0.0849          & 0.0550          & 0.0274          & 0.1300          & 0.0897          & 0.0652          & 0.0399          & 0.0428          & 0.0234          \\
                               &                                    & \Checkmark                       & 0.0324          & 0.0160          & 0.1252          & 0.0844          & 0.0567          & 0.0272          & 0.1242          & 0.0868          & 0.0654          & 0.0399          & 0.0437          & 0.0238          \\ \cline{1-3}
\Checkmark                     & \Checkmark                         &                                  & {\ul 0.0498}    & {\ul 0.0247}    & {\ul 0.1455}    & {\ul 0.0938}    & 0.0583          & 0.0295          & {\ul 0.1294}    & 0.0872          & {\ul 0.0699}    & {\ul 0.0425}    & 0.0476          & 0.0262          \\
\Checkmark                     &                                    & \Checkmark                       & 0.0450          & 0.0240          & 0.1423          & 0.0920          & {\ul 0.0592}    & 0.0286          & 0.1274          & 0.0898          & 0.0689          & 0.0409          & {\ul 0.0488}    & {\ul 0.0275}    \\
                               & \Checkmark                         & \Checkmark                       & 0.0401          & 0.0234          & 0.1378          & 0.0915          & 0.0587          & {\ul 0.0300}    & 0.1288         & {\ul 0.0913}    & 0.0685          & 0.0412          & 0.0469          & 0.0259          \\ \cline{1-3}
\Checkmark                     & \Checkmark                         & \Checkmark                       & \textbf{0.0550} & \textbf{0.0291} & \textbf{0.1510} & \textbf{0.0979} & \textbf{0.0620} & \textbf{0.0339} & \textbf{0.1359} & \textbf{0.0931} & \textbf{0.0740} & \textbf{0.0455} & \textbf{0.0533} & \textbf{0.0309} \\ \hline
\end{tabular}
\caption{Ablation of key strategies on three datasets, best results are marked in bold,
second best results underlined.}
\label{tab:allabl}
\end{table*}

\subsection{Evaluation Metrics}

Recall measures whether the ground-truth item (or explanation) appears in the top-k predictions. NDCG (Normalized Discounted Cumulative Gain) \cite{ndcg} further considers the position of correct predictions, assigning higher rewards to items ranked higher in the prediction list. Higher Recall and NDCG \cite{ndcg} values indicate better recommendation and explanation performance.

\paragraph{Recall@K.}
For a user $u$, let $G_u$ be the set of relevant items in the test set
and $\widehat{R}_u^{(K)}$ the top-$K$ recommendations.
The per-user recall is
\begin{equation}
  \mathrm{Recall@K}(u)=
  \frac{\lvert\,\widehat{R}_u^{(K)} \cap G_u\,\rvert}{\lvert G_u\rvert}.
\end{equation}
Averaging over all test users $\mathcal{U}$ yields
\begin{equation}
  \mathrm{Recall@K}=
  \frac{1}{\lvert\mathcal{U}\rvert}
  \sum_{u\in\mathcal{U}}
  \frac{\lvert\,\widehat{R}_u^{(K)} \cap G_u\,\rvert}
       {\lvert G_u\rvert}.
\end{equation}
When $\lvert G_u\rvert=1$ (next-item prediction), Recall@K
reduces to Hit@K.

\paragraph{NDCG@K.}
Given the ranked list, let $r_{u,i}$ denote the relevance (1/0) of the
item at position $i$ for user $u$.
The discounted cumulative gain is
\begin{equation}
  \mathrm{DCG}_u@K=
  \sum_{i=1}^{K}\frac{r_{u,i}}{\log_2(i+1)}.
\end{equation}
The ideal DCG, obtained by perfect ranking, is
\begin{equation}
  \mathrm{IDCG}_u@K=
  \sum_{i=1}^{\min(K,\lvert G_u\rvert)}
  \frac{1}{\log_2(i+1)}.
\end{equation}
We normalise DCG to
\begin{equation}
  \mathrm{NDCG}_u@K=
  \frac{\mathrm{DCG}_u@K}{\mathrm{IDCG}_u@K}\in[0,1],
\end{equation}
and report the average over all users:
\begin{equation}
  \mathrm{NDCG@K}=
  \frac{1}{\lvert\mathcal{U}\rvert}
  \sum_{u\in\mathcal{U}}\mathrm{NDCG}_u@K.
\end{equation}

\begin{figure}[!t]
    \centering
    \includegraphics[width=1\linewidth]{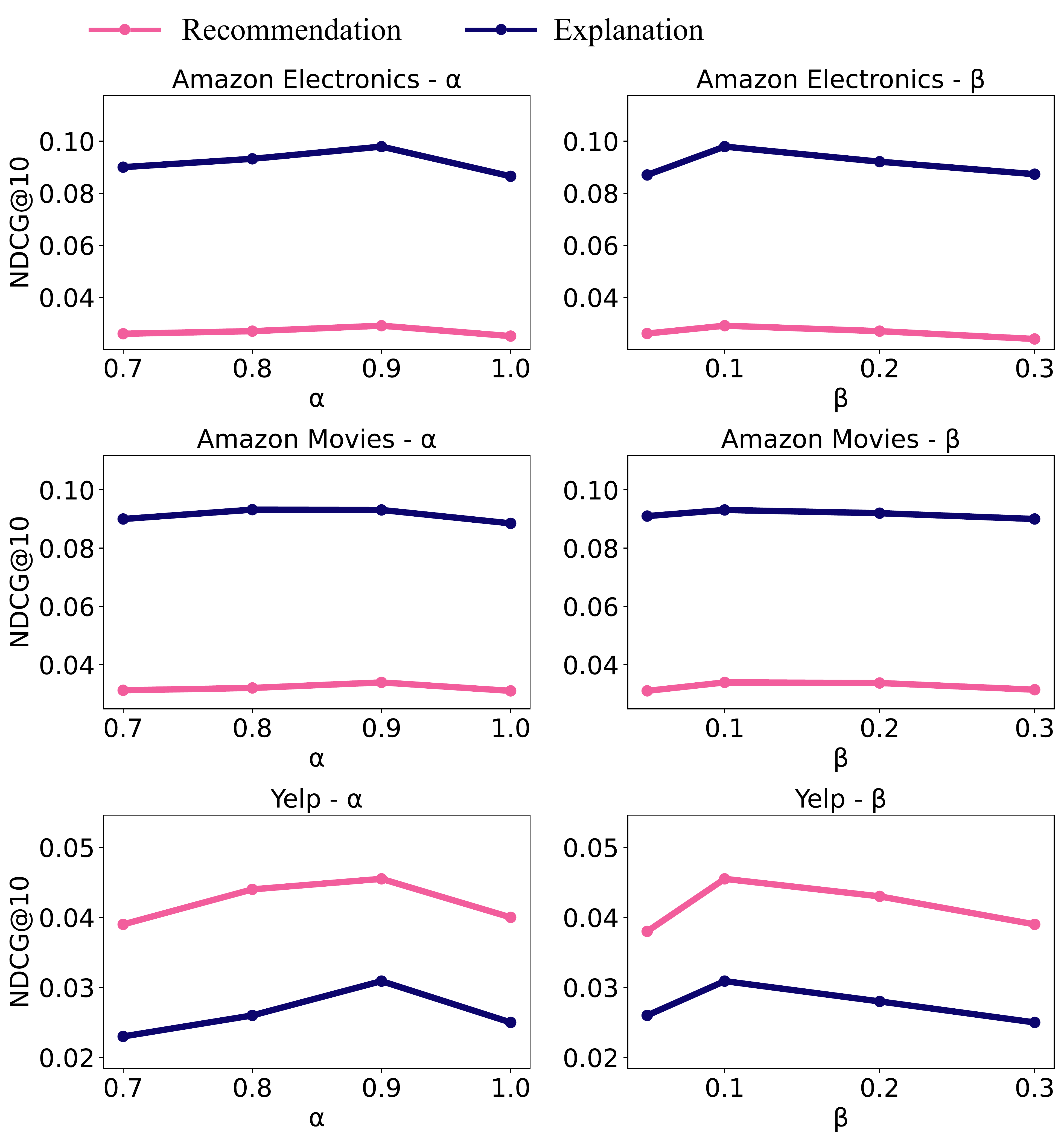}
    \caption{Impact of hyperparameters across three datasets. Results show that when set $\alpha=0.9$ and set $\beta=0.1$, the proposed model achieve the best performance.}
    \label{fig:hyp_sensitivity}
\end{figure}

\subsection{Hyperparameter Sensitivity Analysis}

\Cref{fig:hyp_sensitivity} plots NDCG@10 on the three datasets while sweeping the two scalar hyperparameters that control feature fusion in \textsc{TME-PSR}. The hyperparameter $\alpha$ determines the degree of fusion between the recommendation base embedding $\mathbf{E}_\mathcal{V}$ and the explanation base embedding $\mathbf{E}_\mathcal{E}$ features both in the recommendation branch and the explanation branch. The hyperparameter $\beta$ controls the contribution of the temporal features $\tilde{\mathbf{E}}^{\text{rec}}$ and $\tilde{\mathbf{E}}^{\text{exp}}$ in the final personalized time-aware embeddings $\mathbf{E}^{\text{rec-time}}$ and $\mathbf{E}^{\text{exp-time}}$. When varying the value of one hyperparameter, we keep the other fixed. With $\beta$ fixed to 0.1, we turn over $\alpha$ in $\{0.7,0.8,0.9,1.0\}$. Increasing $\alpha$ from 0.7 to 0.9 consistently improves both recommendation performance and explanation performance. Then performances drops obviously when $\alpha=1.0$, indicating the recommendation embedding and the explanation embedding are not fused. With $\alpha$ fixed to 0.9, we turn over $\beta$ in $\{0.05,0.1,0.2,0.3\}$. We observed that $\beta=0.1$ yields the best results. Performance drop significantly when it goes beyond $0.1$. On the other hand, when the $\beta$ is too small, like $0.05$, will cause the temporal feature to have too little impact to work.

\begin{figure}[t]
    \centering
    \includegraphics[width=1\linewidth]{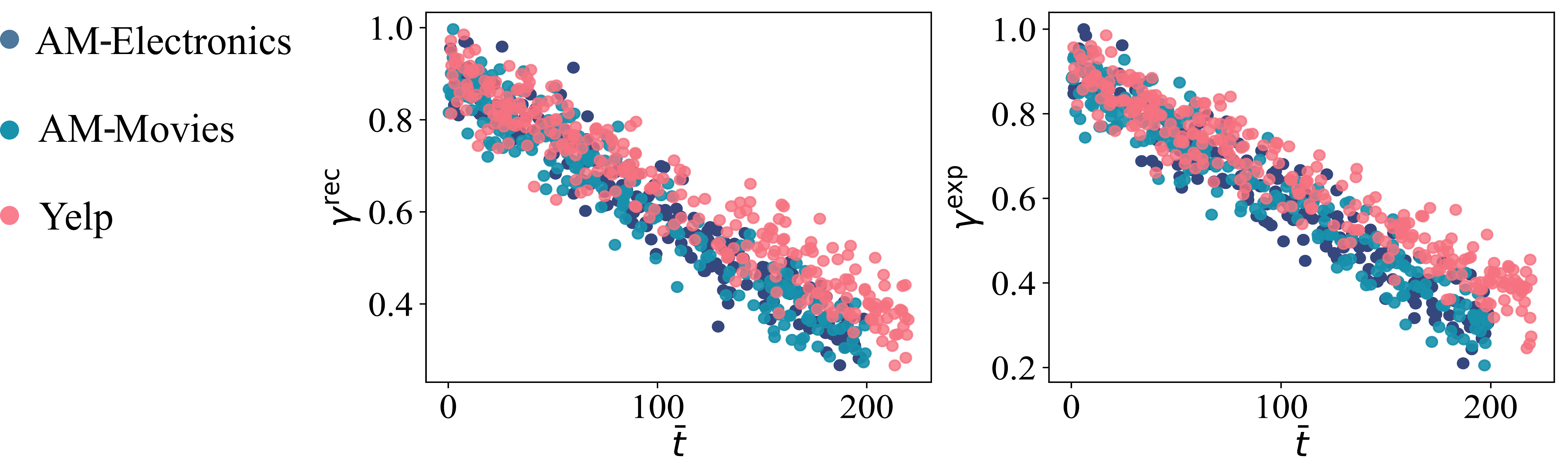}
    \caption{Relationship between the gating parameters $\gamma^{\text{rec}}$ and the average time interval $\bar{t}$ (left) and relationship between the gating parameters $\gamma^{\text{exp}}$ and $\bar{t}$ (right). Each point represents a single user sequence. Results show that as $\bar{t}$ increases, the two gating parameters decrease linearly.}
    \label{fig:gamma_rel}
\end{figure}

\section{Complete comparisons with baseline models}
We compare the proposed TME-PSR with the origin models without SCEMIM. As \cref{tab:overall_onlyrec} shows, the proposed method still achieved significant improvements in recommendation metrics across all three datasets.

\section{Complete ablation study}
We evaluate every non-empty combination of the three personalization strategies on the three datasets, as \cref{tab:allabl} shows. The main findings are entirely consistent with the ablation study described in sec 4. In summary, while the relative contribution of each component varies with the dataset characteristics, the complete TME-PSR configuration reliably delivers the strongest recommendation and explanation quality across all settings.

\section{Time-aware Personalization Details}

We further explore the relationships between the two gating parameters $\gamma^{\text{rec}}$ and $\gamma^{\text{exp}}$ and the average time interval $\bar{t}$ of each user sequence. Specifically, for each user sequence $S_u$, we compute the average time interval $\bar{t}=\frac{1}{n-1}\sum^n_{i=2}(t_i-t_{i-1})$. \Cref{fig:gamma_rel} visualizes 2D points $(\bar{t},\gamma^{\text{rec}})$ and 2D points $(\bar{t},\gamma^{\text{exp}})$ for different user sequences. Results show that as the average time interval $\bar{t}$ increases, the two gating parameters $\gamma^{\text{rec}}$ and $\gamma^{\text{exp}}$ decrease linearly. That is, sequences with short intervals tend to have larger $\gamma^{\text{rec}}$ and $\gamma^{\text{exp}}$. This aligns with our design intuition: sequences with short intervals reflect recent, dense behaviors and thus benefit more from short-term temporal modeling, because large $\gamma^{\text{rec}}$ and $\gamma^{\text{exp}}$ mean a greater contribution of short-term temporal features to $\tilde{\mathbf{E}}^{\text{rec}}$ and $\tilde{\mathbf{E}}^{\text{exp}}$. In contrast, sequences with long intervals rely more on long-term temporal modeling.

\section{Complexity analysis of multihead LRU}

We analyze a single-layer LRU for full-length forward with recursive parallelization. For one head with hidden size $d$, the time complexity is $O(\log(n)\cdot d^2)$ \cite{lrurec} for a sequence of length $n$. We ignore constant factors and linear-time terms due to normalization/residual wiring and do not account for additional FFN/projection costs here. Then, the time complexity of serially computing all heads is $O(H\cdot\log(n)\cdot (d/H)^2) = O(\log(n)\cdot\frac{d^2}{H})$. The time complexity of ideal parallel computing of all heads is $O(\log(n)\cdot(\frac{d}{H})^2)$. With per-head $\mathbf{W}_h,\mathbf{U}_h\in \mathbb{R}^{\frac{d}{H}\times \frac{d}{H}}$, the space complexity of per head is $O((\frac{d}{H})^2)$ and the total space complexity of the entire architecture is $O(\frac{d^2}{H})$, which is inversely proportional to the number of heads $H$. In comparison, typical transformer-based recommenders has a time complexity of $O(n^2d+nd^2)$ and a space complexity of $O(d^2)$ and typical RNN-based recommenders has a time complexity of $O(nd^2)$ and a space complexity of $O(d^2)$. 


\begin{figure}[t]
    \centering
    \includegraphics[width=1\linewidth]{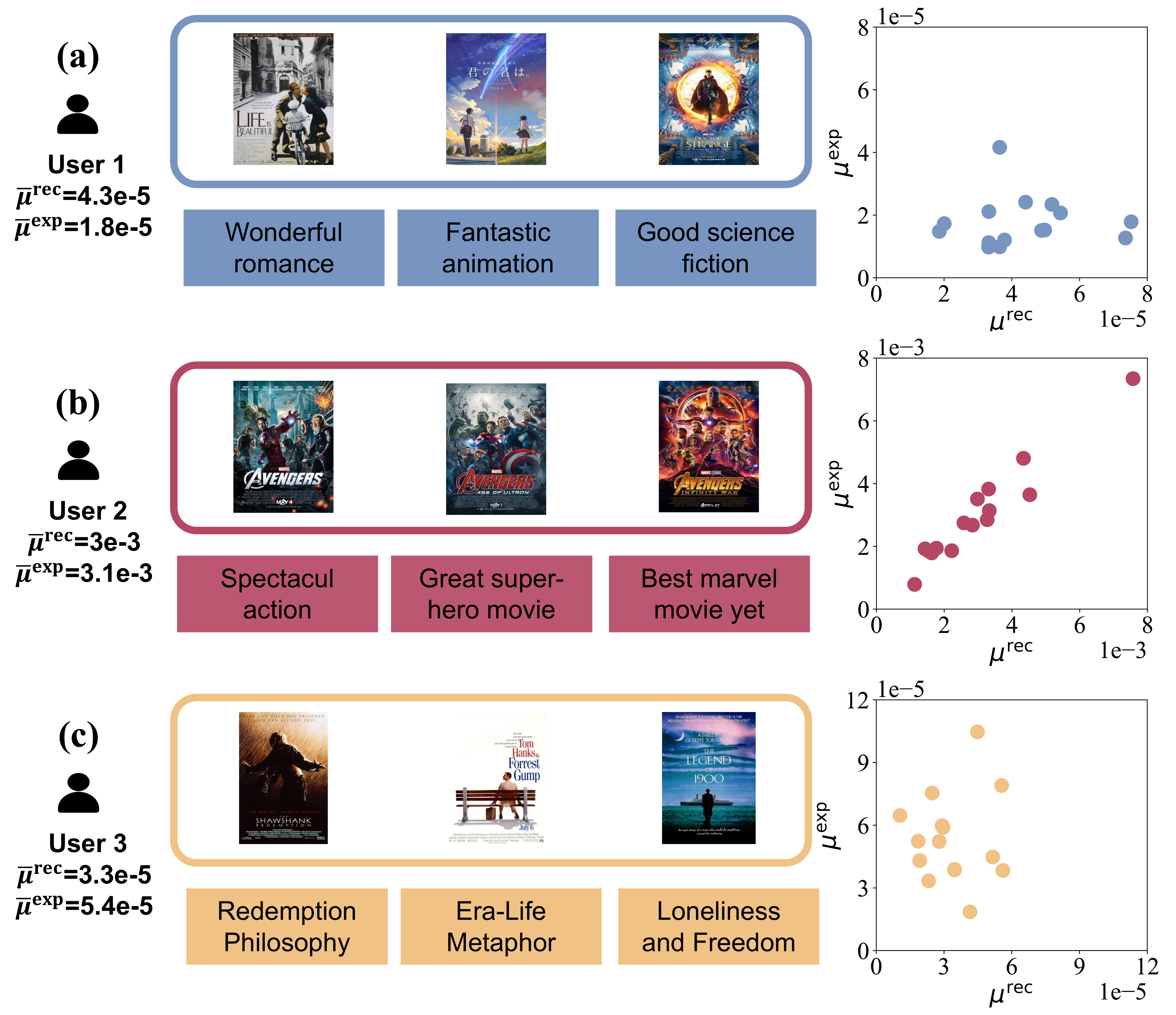}
    \caption{Examples of three representative user types, correspond to the three clusters in Figure 7(b).}
    \label{fig:mu_case}
\end{figure}

\section{Examples of three representative user}
Fig. ~\ref{fig:mu_dis}(a) shows that all distributions are highly left-skewed and concentrated near zero, indicating that most sequences are assigned low semantic alignment strengths between recommendations and explanations in both the recommendation task and the explanation task. For example, User 1 in \cref{fig:mu_case}(a) and User 3 in \cref{fig:mu_case}(c) have low values of $\bar{\mu}^{\text{rec}}$ and $\bar{\mu}^{\text{exp}}$. Meanwhile, the long tail of the distribution ensures a portion of sequences still receive strong semantic alignment signals when needed. For example, User 2 in \cref{fig:mu_case}(b) has high values of $\bar{\mu}^{\text{rec}}$ and $\bar{\mu}^{\text{exp}}$. 

Fig. ~\ref{fig:mu_dis}(b) shows that user-level $\tilde{\mu}^{\text{rec}}_u$ and $\tilde{\mu}^{\text{exp}}_u$ exhibit three well-separated clusters: (1) The user shown in \cref{fig:mu_case}(b) belongs to the balanced group, this type of users has similar semantic alignment requirements in the recommendation task and the explanation task. (2) The user shown in \cref{fig:mu_case}(a) belongs to the recommendation-dominant group, this type of users has higher semantic alignment requirements in the recommendation task. (3) The user shown in \cref{fig:mu_case}(c) belongs to the explanation-dominant group, this type of users has higher semantic alignment requirements in the explanation task.

\end{document}